\documentclass[aps,prx,superscriptaddress,amsmath,amssymb,twocolumn,a4paper,floatfix]{revtex4-2}

\usepackage{graphicx}
\usepackage{dcolumn}
\usepackage{bm}
\usepackage{braket}
\usepackage{hyperref}
\usepackage{xcolor}

\begin{document}

\title{Collective theory for an interacting solid in a single-mode cavity}
	
\author{Katharina Lenk}
\affiliation{Department of Physics, University of Erlangen-N\"urnberg, 91058 Erlangen, Germany}
\author{Jiajun Li}
\affiliation{Department of Physics, University of Fribourg, 1700 Fribourg Switzerland}
\affiliation{Paul Scherrer Institute, Condensed Matter Theory, PSI Villigen, Switzerland}
\author{Philipp Werner}
\affiliation{Department of Physics, University of Fribourg, 1700 Fribourg Switzerland}
\author{Martin Eckstein}
\affiliation{Department of Physics, University of Erlangen-N\"urnberg, 91058 Erlangen, Germany}

\date{\today}

\begin{abstract}
We investigate the control of interacting matter through strong coupling to a single electromagnetic mode, such as the photon mode in a Fabry-Perot or split-ring cavity. For this purpose, we analyze the exact effective theory for the collective light-matter hybrid modes of a generic system  of $N$  transition dipoles  within an interacting solid. The approach allows to predict properties of the coupled light-matter system from the nonlinear response functions of the uncoupled matter ``outside the cavity''. The limit of  large $N$ corresponds to a conventional macroscopic description based on the polarizability of matter. In this limit, the cavity does not affect the static ferroelectric response. Corrections, which are needed to understand finite size systems and to obtain the nonlinear light-matter response, can be obtained from the non-linear susceptibilities of the matter outside the cavity. The theory is benchmarked for the Dicke model, and for a quantum Ising model which serves as a minimal mean-field model for a quantum paraelectric material like SrTiO\textsubscript{3}.
\end{abstract}


\maketitle

\section{Introduction}

Enhancing the light-matter coupling in cavities provides an intriguing route to control properties of matter, from chemical reactions to transport and thermodynamic phase transitions \cite{Schachenmayer2015, Feist2015, ebbesen2016, Schlawin2022, thomas2021, appugliese2021}. 
Order parameters which couple linearly to the electromagnetic field, such as ferroelectricity, incommensurate charge density waves, or exciton condensates, appear most suitable in this context, but the possible mechanisms are not well understood in many cases.  An early proposal for cavity-induced ferroelectricity in an ensemble of independent emitters is the equilibrium Dicke superradiant phase \cite{Hepp1973,Wang1973}. The latter is not stable when the proper gauge-invariant light-matter coupling is taken into account \cite{Rzazevski1975}, but phases like ferroelectricity can already be induced by intrinsic interactions in matter, and it remains a valid question to what extent the coupling to cavity modes can enhance such collective behavior \cite{Keeling2007, Bernardis2018, Mazza2019, Ashida2020, Lenk2020, Latini2021}.

In this work we analyze the effect of quantum light on interacting solids in a single-mode cavity. The single-mode cavity setting has been frequently studied in theoretical works as a promising platform to engineer periodic electronic systems \cite{Kiffner2019, Kiffner2019b, Wang2019, Sentef2020, Li2020, Latini2021}. It can be realized, e.g., by a Fabry Perot cavity, or by a near-field cavity such as a split-ring resonator \cite{Maissen2014}  where the electric field of a given mode with frequency $\Omega$ is confined to a volume $V$ much smaller than $\lambda^3=(\Omega/c)^3$. The mode volume therefore fits only a finite number $N$ of atoms, which however can still be macroscopic ($N\sim10^{10}$) even for a $\mu$m size cavity. There are then two relevant energy scales which should be compared to the cavity frequency $\hbar\Omega$ and the bare energies in matter, such as a level splitting $\Delta$: (i) The single-particle coupling $g_1^2$, which determines the interaction between a single photon and a single transition dipole in the solid, and (ii) the collective coupling $g_n^2=Ng_1^2$, which quantifies the hybridization of a collective excitation in matter with the cavity mode. Although ultra-strong single-particle coupling $g_1^2\sim\Delta$ has been achieved in experiment \cite{Kockum2019}, this regime is restricted to few emitters and cannot provide a generic route to control condensed matter phases: More precisely, the single-particle coupling is inversely proportional to the mode volume, $g_1^2\sim1/V$, but the maximum number of emitters in a given volume $V$ is limited by the density $n=N/V$. For a cavity which is large enough to accommodate a macroscopic number of atoms, the single-particle coupling is therefore  small  even when the cavity geometry compresses the light mode deep into the subwavelength regime, while the maximum collective coupling $g_n^2$ is proportional to the density and thus finite for $N\to\infty$.

In a dense solid, this collective coupling can reach electronic energies (eV) and can easily be comparable or larger than other bare energy scales. This raises the fundamental question whether collective strong coupling to a single mode can affect static properties of matter, even when the single-particle coupling is weak. So called ``no-go'' theorems exclude condensation of a single homogeneous photon mode  in the thermodynamic limit under quite general assumptions \cite{Andolina2019, Andolina2020, Andolina2021}, but beyond that theoretical proposals still suggest the possibility of cavity-induced ferroelectricity in a single mode cavity \cite{Latini2021}. For finite systems, intriguing effects can arise  due to single-mode coupling \cite{Kiffner2019, Kiffner2019b, Wang2019, Sentef2020, Li2020}, but their fate in the limit $N\to\infty$ is less clear. To shed light on these issues, one can aim to develop a theory that describes directly the collective hybrid light-matter response. For example,  this approach has been followed in Ref.~\onlinecite{Latini2021} by deriving a collective Hamiltonian for the zone-centered IR-active phonons in SrTiO$_3$. Here we analyze exact properties of the collective theory at arbitrary collective coupling, for the general case of a solid in which the cavity mode couples to $N$ dipoles (which represent, for example, the displacements of ions within a unit cell, or electronic transition dipoles), allowing for an arbitrary direct interaction between the dipoles, and  between the dipoles and other degrees of freedom in the solid. The aim is to quantify the properties of matter in the cavity exactly in terms of the nonlinear response functions of the material which is not coupled to the cavity mode (``matter outside the cavity''), so that the approach can serve both as a  starting point for computational and phenomenological studies.  

Before presenting this approach, let us emphasize that the single-mode cavity, with  mesoscopic length scales, should be clearly distinguished from an extended cavity or a waveguide, such as a coplanar cavity which confines the light only in one direction \cite{Jarc2021}. In such cavities matter can  easily be taken to the thermodynamic limit, in which case the single-mode coupling vanishes, but there is a continuum of modes with different in-plane momenta which can have a non-vanishing combined effect on matter \cite{Sentef2018b, Schlawin2019, li2022, Ashida2020, Rokaj2022}. The challenge is  to ensure that the coupling affects a broad momentum range, whereas in free space only photons with momenta much smaller than the extent of the Brillouin zone are relevant for the low energy physics. For example, ferroelectricity can be influenced in a coplanar cavity if the coupling to slowly propagating surface plasmon modes at the cavity-matter interface is taken into account \cite{Ashida2020}. 

This work is structured as follows. In Sec.~\ref{sec:coll} we summarize the main ideas behind the collective approach. Details of the derivation are given in the Appendices. In Sec.~\ref{Sec:dicke}, we illustrate the collective theory for the Dicke model, where  the matter is given by isolated dipoles which interact only via the cavity mode. The cavity cannot induce a phase transition in this model \cite{Rzazevski1975}, but it does affect the static properties for $N<\infty$. Even the lowest order collective theory can reproduce the finite-$N$ corrections accurately, also in the  ultra-strong collective coupling regime. In Sec.~\ref{Sec:all2all}, we then analyze a minimal model for a so-called quantum paraelectric, where we find an enhancement of the static polarizability (again for finite $N$). Finally, Sec.~\ref{Sec:conclusion} provides a summary and conclusion.

\section{Collective theory}
\label{sec:coll}

\subsection{Setting}

We consider a generic molecular solid consisting of $N$ units (e.g., polarizable atoms or molecules). The Hamiltonian $H_{\rm mat}$ describing the isolated matter may include an arbitrary direct interaction between the units, such as dipolar interactions arising from the longitudinal electromagnetic fields (Coulomb interaction), or electron-lattice interactions which can drive a ferroelectric ordering. The transverse electromagnetic field is described by a single-mode with bare Hamiltonian $H_{\rm field}=\Omega \hat a^\dagger \hat a$  ($\hbar=1$).  With a constant electric field confined in the mode volume $V$ and a background dielectric constant $\epsilon$, the electric field operator is $\hat E = \sqrt{\frac{\Omega}{V \epsilon}}\hat X$, where $\hat X=(a^\dagger+\hat a)/\sqrt{2}$ is the field quadrature.  For the light-matter interaction we specifically assume a dipolar transition operator $ed\hat p_r$ at each unit $r$, with the elementary electric charge $e$, a length scale $d$, and a dimensionless operator $\hat p_r$ (e.g., this could be the $\hat \sigma_x$ operator when the units are described by two level systems). The linear coupling to the electric field then implies the interaction term 
\begin{equation}
\label{hep}
	\hat{H}_{\rm Ep}= \sqrt{\Omega} g_1 \hat X \hat P,
\end{equation}
where $\hat P=\sum_{r}\hat{p}_{r}$ is the total dipole,  and $g_1=\sqrt{\frac{e^2d^2}{V \epsilon}}$ the single-particle coupling constant. In addition we must add the term $\hat{H}_{\rm pp}=\frac{1}{2}g_1^2 \hat P^2$, which completes the light-matter interaction into a positive definite form as required from the derivation of the light matter Hamiltonian in the dipolar form within the Coulomb gauge \cite{Di-Stefano2019, Bernardis2018, Jiajun2020, Schaefer2020}. Note that $\hat H_{\rm pp}$ is not the electrostatic dipolar interaction, which instead arises from the longitudinal fields.  The positive definiteness of the Hamiltonian is evident by writing 
\begin{align}
\label{square}
H_{\rm pp}+H_{\rm field}+H_{\rm Ep}= \Omega \hat b^\dagger \hat b, 
\end{align}
with $\hat b  = \hat a + g_1\hat P/\sqrt{2\Omega}$. The relevance of the term $H_{\rm pp}$, analogous to the relevance of the diamagnetic term in the $p\cdot A$ representation of the light-matter coupling, has been emphasized in various contexts, including the stability of matter \cite{Schaefer2020} and light-induced ferroelectricity \cite{Lenk2020}. We shall see that this property is also important for the results below.

To solve this model, we use field-theoretical techniques based on the imaginary-time path-integral formalism. The detailed calculations are given in the appendices, and we summarize in the main text the results and the idea behind the derivation. Our aim is to obtain both the response function $D(\tau) = -\langle T_\tau \hat a(\tau) \hat a^\dagger (0)\rangle$ of the photon field, and the collective response function $\chi(\tau) = \frac{1}{N} \langle T_\tau \hat P(\tau) \hat P (0)\rangle$ of matter. Within the imaginary time formalism, correlation functions will be expressed as functions of imaginary time $\tau$ or  Matsubara frequency $i\nu_m$; the physical (retarded) response functions can then be obtained by replacing $i\nu_m \to \omega +i0$. In particular, $\chi(\omega+i0)$ gives the linear response of $\frac{1}{N}\langle \hat P(t)\rangle$ to a force $h(t)$ which couples through a term $-h(t)\hat P$ in the Hamiltonian, and is $\mathcal{O}(1)$ for large $N$.

\subsection{Light-induced interactions} 

To investigate the properties of the system in thermal equilibrium, one can first exactly integrate out the photon fields, which gives rise to the frequency-dependent (retarded) all-to-all induced interaction (App.~\ref{App:lightinduced})
\begin{align}
\label{vinddd}
V_{\rm ind}(i\nu_m) 
&=  
g_n^2 
\frac{\nu_m^2}{\nu_m^2+\Omega^2}.
\end{align}
This interaction, with the collective coupling $g_n^2 =Ng_1^2$ is normalized such that the individual interaction for a pair of atoms is $V_{\rm ind}/N$ [c.f.~\eqref{sindgehe} in the appendix]. It combines both the direct interaction  coming from $H_{\rm pp}$ and the photon-mediated terms. These terms exactly cancel in the static limit $\nu_m=0$ as a consequence of the complete square form \eqref{square} of the Hamiltonian, which implies that a static shift of $\hat P$ can be gauged into a change of the field operators $\hat a$ and therefore does not lead to a change of the energy. 

The exact relation between the photon propagator and the response of the matter is given by (App.~\ref{app:exactrelations01})
\begin{align}
\label{dinteracting}
D(i\nu_m)=
D_0(i\nu_m)
-
D_0(i\nu_m)
\frac{\Omega g_n^2}{2}
\chi(i\nu_m)
D_0(i\nu_m),
\end{align}
where $D_0(i\nu_m)=(i\nu_m-\Omega)^{-1}$ is the noninteracting photon propagator. The real-frequency poles of $D$ give the energies of the hybrid light-matter modes. As expected, the hybridization of the bare photon and matter, and thus properties like the Rabi splitting between the light and matter modes,  is set by the collective coupling $g_n^2$. 

The effect of the light-matter interaction on the matter is more involved. Within a heuristic mean-field treatment, one would simply replace the response of $p=\frac{1}{N}\langle \hat P\rangle$ in the cavity to a force $h$ (coupling $-h\hat P$ in the Hamiltonian) by the bare response $\chi_{\rm mat}$ to both  $h$ and the induced mean-field force $-V_{\rm ind}\,p$; i.e., $p=\chi_{\rm mf}h \approx \chi_{\rm mat} (h-V_{\rm ind}\,p)$. This leads to the standard RPA equation
\begin{align}
\chi_{\rm mf}(i\nu_m) = \frac{\chi_{\rm mat}(i\nu_m)}{1+\chi_{\rm mat}(i\nu_m)V_{\rm ind}(i\nu_m)},
\label{chimf}
\end{align}
where $\chi_{\rm mat}$ is the response of matter not coupled to the cavity. This theory is in essence a macroscopic description in 
which a medium with a frequency-dependent polarizability $\chi_{\rm mat}$ is collectively coupled to the cavity field.  Because $V_{\rm ind}(i\nu_m=0)=0$, Eq.~\eqref{chimf} shows that the static response is left unchanged by the cavity within the mean-field description, which would also rule out an effect of the cavity on thermodynamic properties and phase transitions. 

\begin{figure}[tbp]
\centerline{\includegraphics[width=0.48\textwidth]{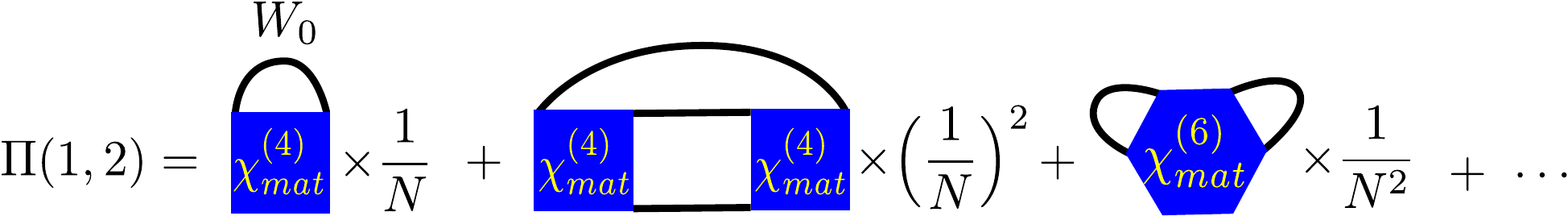}}
\caption{Diagrams for the self energy $\Pi$. The lines are given by the ``noninteracting'' $\varphi$-propagator \eqref{W0line}, while the interaction vertices are the connected correlation functions of the matter, with the corresponding scaling with $N$, i.e. the $n$-point interaction vertex is $\chi_{\rm mat}^{(n)}/N^{\frac{n-2}{2}}$.}
\label{figdiagrams}
\end{figure}

\subsection{Collective Theory} 
\label{subseccoll}

To go beyond the mean-field description, we write down a theory for the collective modes (App.~\ref{App:HubStrato}). It is convenient to perform a Hubbard-Stratonovich transformation and choose as a basic collective field the dual variable $\varphi$ of $\hat P$, which mediates the interaction $V_{\rm ind}$. The full imaginary time action in terms of this field is given by $S=S_{\rm mat} + \frac{1}{2} \varphi_{j} (V_{\rm ind}^{-1})_{jj'} \varphi_{j'}+ \frac{i}{\sqrt{N}} \varphi_j P_{j}$; here, the fields are written in terms of discrete time indices $j$, 
and a summation over repeated indices with a factor $\Delta\tau$ is implied, $a_jb_j\equiv\int d\tau a(\tau) b(\tau)$. 
The action is defined such that upon integrating out the field $\varphi$, the interaction $V_{\rm ind}$ within the matter system is recovered. In turn, integrating out the matter will give an interacting theory for $\varphi$, with $\varphi^4$  and higher order interactions which can be expressed 
exactly in terms of the $n$-th order nonlinear response functions $\chi_{\rm mat}^{(n)}$ of the matter which is not coupled to the cavity. More specifically, the exact action is given by
\begin{align}
\label{collectiveaction}
S[\varphi] = \frac{1}{2} \varphi_{j} (W_0^{-1})_{jj'}\varphi_{j'} + \sum_{n=4,6,...}^\infty N^{1-\frac{n}{2}} S_{\rm int}^{(n)},
\end{align}
where 
\begin{align}
\label{W0line}
W_{0} (i\nu_m) = \frac{V_{\rm ind}(i\nu_m)}{1+\chi_{\rm mat}(i\nu_m) V_{\rm ind}(i\nu_m)}, 
\end{align}
and  $S_{\rm int}^{(n)}=\frac{1}{n!} \varphi_{j_1}\cdots \varphi_{j_n}  \chi^{(n)}_{\rm mat}(j_1\cdots j_{n})$ with the connected correlation functions  $\chi^{(n)}_{\rm mat}(j_1,j_2,...)=\frac{1}{N}\langle T_\tau P(\tau_1) P(\tau_2) \cdots \rangle^{\rm con}_{\rm mat}$ of the isolated matter. The latter are in essence the nonlinear response functions, which are $\mathcal{O}(1)$ for large $N$. Note that only even order interaction terms are present as we assume parity symmetry in the problem. 

An action of this form, where $n$-point interactions between an auxiliary Hubbard-Stratonovich field $\varphi$ are given in terms of physical $n$-point correlation functions, naturally appears when interacting degrees of freedom are integrated out to obtain the action for $\varphi$. Examples where such an action is used as a starting point for a computational approach are the dual fermion \cite{Rubtsov2009} and dual boson \cite{Rubtsov2012} theories. Also in the present case, the theory for the $\varphi$-field can be used directly to compute properties of matter. The corrections for the full propagator $W(\tau) = \langle T_\tau \varphi(\tau)\varphi(0)\rangle$ can be defined in terms of a  self-energy $\Pi$, which determines $W$ via the Dyson equation $W^{-1}=W_{0}^{-1}-\Pi$. Following the exact relations between matter and $\varphi$, one has (App.~\ref{App:parturbation})
\begin{align}
\chi(i\nu_m)= \frac{\chi_{\rm mat}(i\nu_m)-\Pi(i\nu_m)}{1+\big[\chi_{\rm mat}(i\nu_m)-\Pi(i\nu_m)\big]V_{\rm ind}(i\nu_m)}.
\label{chifinal}
\end{align}
In the static limit ($i\nu_m=0$), we have $V_{\rm ind}(0)=0$, so that only the numerator remains in this expression. This implies that $-\Pi(0)$ is the cavity correction to the static susceptibility. 

All what remains is therefore to quantify the interaction correction $\Pi$ from the interacting theory \eqref{collectiveaction}. The zeroth order ($\Pi=0$) is just the heuristic mean-field approach mentioned in Eq.~\eqref{chimf}, and corrections can be calculated with standard diagrammatic rules. The lowest-order diagrams for $\Pi$ are shown in Fig.~\ref{figdiagrams}, where lines represent the free propagator $W_{0}$ [Eq.\eqref{W0line}], and polygons the interaction vertices. To understand the order of magnitude of these diagrams we consider the following limits: (i) For few emitters ($N\sim1$),  the interaction vertices are of order $1$, but with $g_n^2\sim g_1^2$ the lines $W_0$ become proportional to the single particle coupling [c.f.~Eq.~\eqref{W0line}]. The perturbation series therefore becomes an expansion in the single-particle coupling, and it is controlled by the number of lines. The leading diagram is the Hartree diagram (the first diagram in the figure). (ii) For many particles ($N\gg 1$) and given $g_n^2$ (which can become comparable or larger than the energy scales of the matter), the lines are not necessarily small, but the interaction vertices become small because of the pre-factor $1/N$. The leading diagram is again the Hartree diagram. (Note that, as discussed in the introduction, $g_n$ and with this also the lines $W_0$ remain finite in the limit where the cavity volume $V$ and $N$ are increased at fixed density $n$).

In summary, the exact response of the matter is given by the central result \eqref{chifinal}, which is the macroscopic theory \eqref{chimf} in which the matter response experiences a correction $\Pi$ from the light-matter interaction. The latter has a nontrivial dependence on the collective coupling (see also the examples below), but it  scales like $1/N$ for large $N$, even for non-perturbative collective coupling $g_n^2$, and it also remains small for small single-particle coupling. Both limits are usually well satisfied experimentally.  The vanishing of $\Pi$ in the thermodynamic limit  ($N\to\infty$ at given $g_n$) implies the absence of any cavity effect on the static response, which is in agreement with no-go theorems \cite{Andolina2019, Andolina2020, Andolina2021} that rule out cavity-induced photon condensation.

For completeness, let us also quote a compact expression for the static ($i\nu_m=0$) limit of the Hartree diagram (see App.~\ref{App:HArtree}),\begin{align}
\Pi(0) =
\frac{1}{2N\beta}
\sum_{m}\ \,W_{0}(i\nu_m)\frac{\partial^2 \chi_{h}(i\nu_m)}{\partial h^2} \Big|_{h=0}.
\label{pistatmaintext}
\end{align}
Here $\chi_{h}$ is the response of the matter outside the cavity in the presence of a nonzero static field $-h\hat{P}$ in the Hamiltonian. The presence of the derivative shows that the diagram is determined in fact by a nonlinear response function. This expression is useful for the numerical and analytical evaluations below. More importantly it highlights two  general properties:  (i) While the macroscopic response \eqref{chimf} is not affected by the cavity at zero frequency, the static correction $\Pi(0)$ does depend on all frequencies due to the nonlinearity, and is therefore nonzero for finite $N$. The nonlinearity of the theory feeds back the cavity effect at finite frequencies onto the static response, which is similar to the role of the nonlinearity in the coplanar cavity case \cite{Ashida2020}. (ii) The matter response  $\chi_h$ is small for frequencies above some large energy scale $E_\text{max}$ set by the matter alone, i.e., $\chi_{h}(i\nu_m) \to 0 $ for $\nu_m=2\pi n/\beta \gg E_\text{max}$. On the other hand, $V_{\rm ind}(i\nu_m)$ is small for frequencies $\nu_m\ll\Omega$  [c.f.~Eq.~\eqref{vinddd}], so that the cavity correction \eqref{pistatmaintext} will be small both in the limit of high cavity frequency  $\Omega\gg E_\text{max}$, and in the high temperature limit $\beta E_\text{max}\ll 1$.

\begin{figure}[tbp]
\centerline{\includegraphics[width=0.5\textwidth]{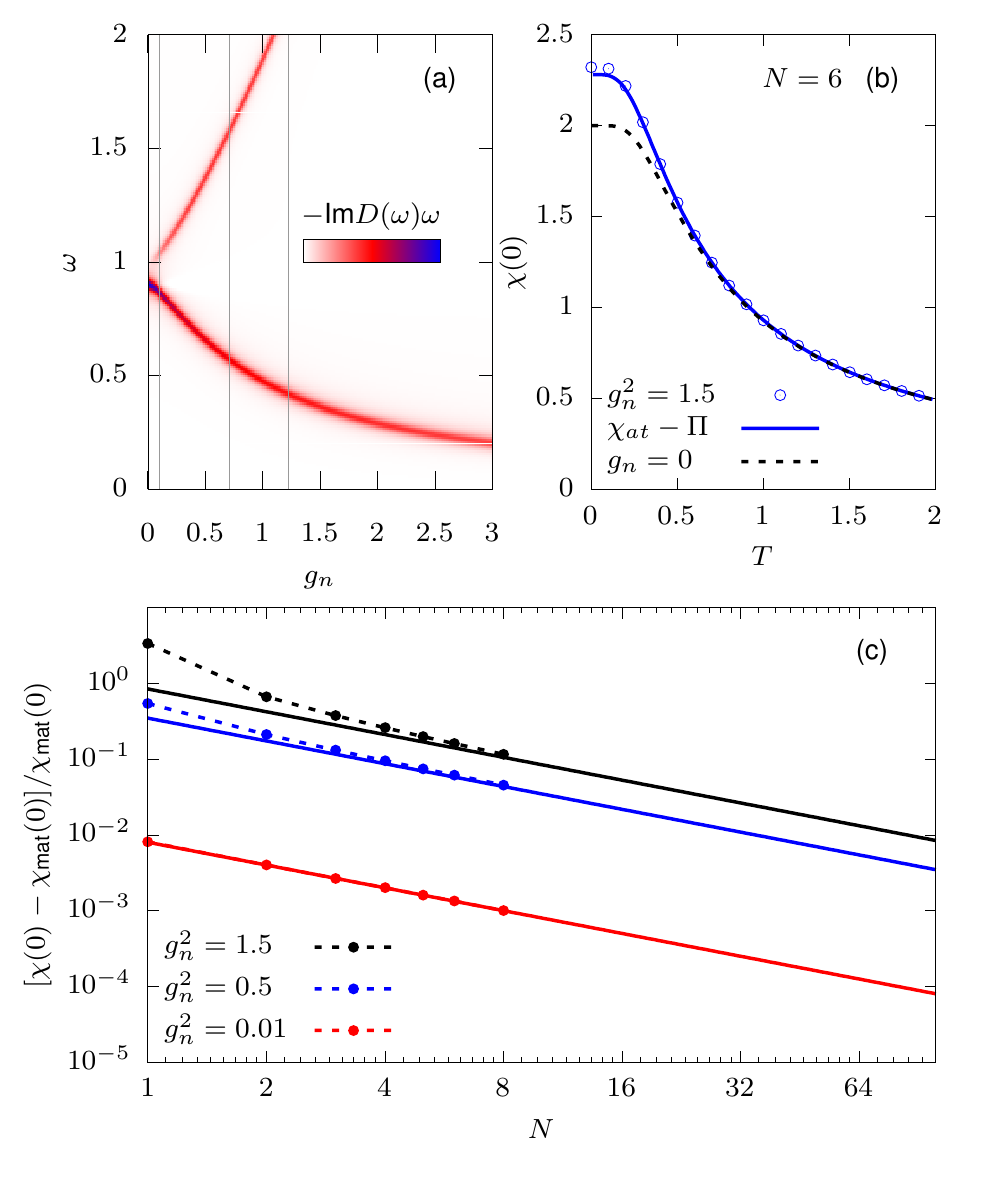}}
\caption{
Dicke model for $\Delta=1$ and $\Omega=0.9$.
(a) Hybrid mode spectrum $-\omega\text{Im}D(\omega+i0)$ at temperature $T=0$ for large $N$, with a Lorentzian broadening of the $\delta$-peaks. The spectrum is plotted as a function of the collective coupling $g_n$, so that the splitting is linear in $g_n$ at weak coupling and $\Delta=\Omega$. (b) Static susceptibility $\chi(0)$ as a function of $T$ for  $N=6$. Symbols: exact diagonalization; solid line: leading order collective theory, Eq.~\eqref{chifinal}; dashed line: exact result for $g_n=0$. (c) Relative difference of $\chi(0)$ to the free value $\chi_{\rm mat}(0)$ at $T=0$, as a function of $N$ for fixed collective coupling $g_n$ (full dots: exact diagonalization, full lines: leading order collective theory). 
The values of $g_n$ correspond to the vertical lines in panel (a).
}
\label{figrabi}
\end{figure}

\section{Evaluation for the Dicke model }
\label{Sec:dicke} 

As a first illustration, we evaluate the $1/N$ correction for the paradigmatic Dicke model, where the only interaction between the atoms is mediated by the field. Here the matter corresponds to $N$ independent two-level atoms with a level splitting $\Delta$,
\begin{align}
H_{\rm mat} = \frac{\Delta}{2}\sum_{r=1}^N \hat \sigma_{z,r}, 
\label{equ:H_mat_Dicke}
\end{align}
and the transition operator in Eq.~\eqref{hep} is $\hat p_{r}=\hat\sigma_{x,r}$ ($\hat \sigma_{\alpha,r}$ are the Pauli matrices acting in the Hilbert space of atom $r$). All energies below will be measured in units of $\Delta$.  The collective response in the limit $N\to\infty$ gives  the well-known Rabi physics. The photon spectrum $-\text{Im} D(\omega+i0)$ [i.e, Eq.~\eqref{chimf} in combination with the photon propagator Eq.~\eqref{dinteracting}]  is shown for illustration in Fig.~\ref{figrabi}(a). One can clearly see the upper and lower polariton branches, i.e., collective modes at the energies which arise from the coupling of two harmonic oscillators with coordinates $X$ and $Q$ and frequencies $\Omega$ and $\Delta$, $H_{mf} = \frac{\Omega}{2}(P_X^2 + (X+g_n/\sqrt{\Omega}Q)^2 + \frac{\Delta}{2} (P_{Q}^2 + Q^2)$.  

We will now analyze the static properties of the system at collective strong coupling, at values of $g_n$ corresponding to the vertical lines in Fig.~\ref{figrabi}a; at the largest values of $g_n$ the Rabi splitting already exceeds the bare energies $\Delta$ and $\Omega$. For the Dicke model, the static correction \eqref{pistatmaintext} to the exact susceptibility Eq.~\eqref{chifinal} can  be evaluated analytically (App.~\eqref{App:Dike}), and the result can be compared to exact diagonalization up to $N\approx10$. For $N\to\infty$, the cavity has no effect on static properties, and $\chi(0)$ is given by the result for an isolated atom, $\chi(0)=\chi_\text{at}(0)=2\tanh(\beta\Delta/2)/\Delta$. Figure~\ref{figrabi}(b) shows the static susceptibility $\chi(0)$ for $N=6$ at $g_n^2=1.5$, comparing exact diagonalization  (symbols) with the leading order collective theory (lines labeled $\chi_\text{at}-\Pi$). One can observe a certain cavity-induced enhancement, which is dominant  for low temperatures in agreement with the argument given above that the cavity effect becomes small in the high-temperature limit. The $1/N$ theory works remarkably well  quantitatively even for these relatively small $N$. Figure \ref{figrabi}(c) systematically analyzes the dependence of the cavity correction $-\Pi(0)$ relative to $\chi_{\rm mat}$ as a function of $N$ for fixed collective coupling, and confirms the fast convergence with $N$ even for strong collective coupling. For weak $g_n^2$, the theory works all the way to $N=1$, as in this case the expansion for $N\sim1$ is controlled by the smallness  of the single-particle coupling $g_1^2$. The behavior is similar for other temperatures and cavity frequencies (not shown here); in particular there is nothing special about the resonance $\Delta=\Omega$. 

\section{Interacting solid}
\label{Sec:all2all}

\begin{figure}[tbp]
	\centerline{\includegraphics[width=0.5\textwidth]{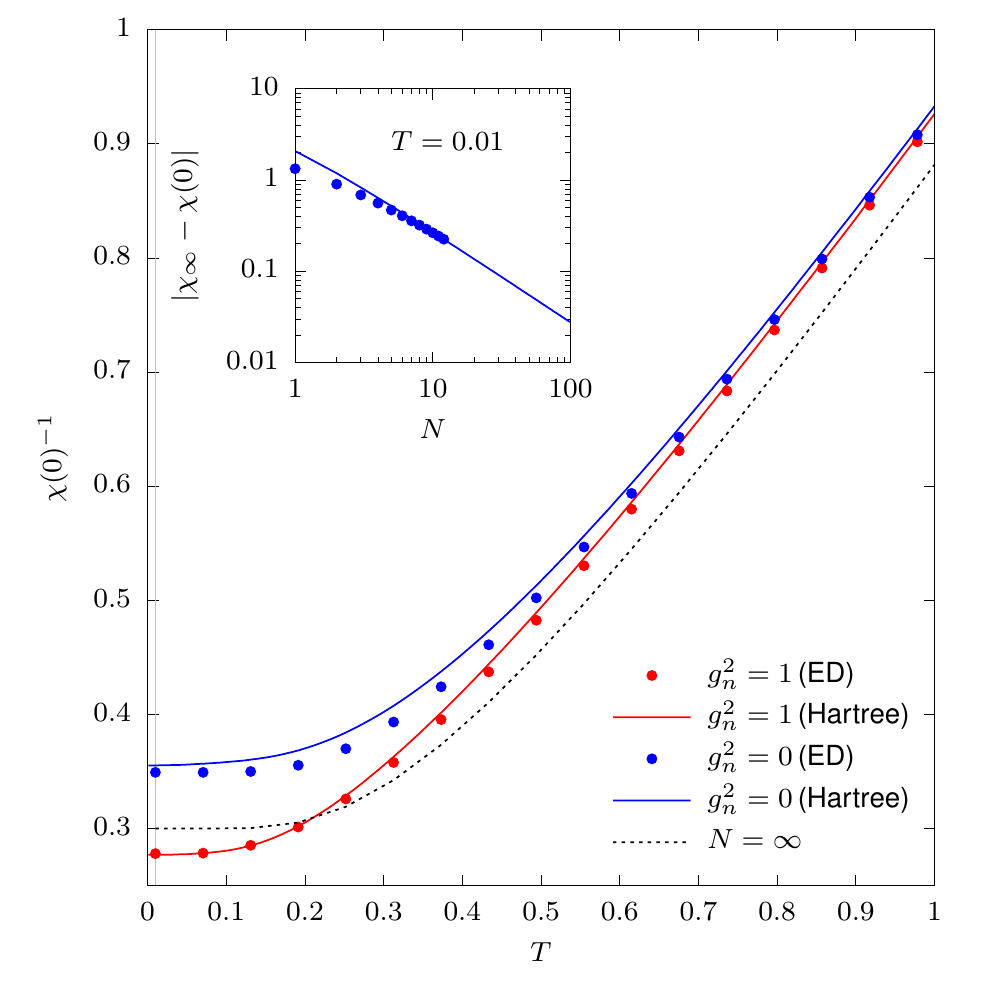}}
	\caption{
	Interacting model for $\Delta=1$ and $\Omega=1$ in the (quantum) paraelectric regime ($\alpha=0.2$). Main panel: Inverse static susceptibility $\chi(0)$ as a function of temperature for $N=5$ emitters,  and collective light-matter coupling $g_n^2=0$ and $g_n^2=1$. The mean-field solution ($N\rightarrow\infty$) is indicated by the black dashed line.   Inset: Difference between the mean-field susceptibility $\chi_{\infty}$ and the static susceptibility $\chi(0)$ at $g_n=0$ and $T=0.01$ (see vertical gray line in the main panel) as a function of $N$. For both panels, symbols and solid lines show results obtained from exact diagonalization (ED) and from the leading $1/N$ theory (Hartree), respectively.}
	\label{fig:ata_temp_alpha0.2}
\end{figure}

\subsection{Model}

In this section, we supplement the model discussed in the previous chapter by a static all-to-all dipole-dipole interaction. The matter Hamiltonian reads
\begin{equation}
\label{equ:Hmat_int}
\hat{H}_{\rm mat}= \frac{\Delta}{2}\sum_{r=1}^N \hat \sigma_{z,r} -\frac{\alpha}{2N}\sum_{r,r'} \hat{\sigma}_{x,r}\hat{\sigma}_{x,r'},
\end{equation}
where the Pauli operators $\hat{\sigma}_{x,r}$ still correspond to the dipole transition matrix elements of the two-level atoms. 
The physics of this model has been studied previously in a wide parameter regime \cite{Bernardis2018}. Here we use it as another demonstration of the  diagrammatic approach in for parameters where the model serves as a mean-field description of a quantum paraelectric material (see below). The all-to-all interaction because this allows for a controlled solution of the model without coupling to the cavity, and therefore an unbiased determination of the cavity effect. In the limit $N\rightarrow\infty$, the dipole-dipole interaction  can drive a second order phase transition to a ferroelectric state, in which the system acquires a non-vanishing macroscopic electric polarization. The classical variant of the model ($\Delta=0$) is an all-to-all Ising model of interacting dipoles, 
for which the phase transition always occurs at a nonzero transition temperature  $T_c$. In the quantum case, however, the phase transition can be inhibited by the term $\Delta$, which allows for a tunneling between the classical configurations. For small values of $\alpha$, the phase transition becomes suppressed even at $T=0$; in this regime, the Hamiltonian \eqref{equ:Hmat_int} therefore  presents a minimal model for a quantum paraelectric \cite{Mueller1979}. More precisely, as discussed in App.~\ref{sec:int_MF},  the static susceptibility in the thermodynamic limit $N\to\infty$ and in the absence of the cavity is given by  the exact mean-field result
\begin{equation}
\chi_{\infty}=\frac{\chi_\text{at}}{1-\alpha \chi_\text{at}}, 
\label{equ:chi_infty}
\end{equation}
where $\chi_{\rm at}(i\nu_m=0,T)=2\tanh(\Delta/2T)/\Delta$ is the static susceptibility of an isolated dipole. The phase transition is therefore determined by the condition
\begin{equation}
 1-\chi_{\rm at}(i\nu_m=0,T_c)\alpha=0,
\end{equation}
which can be satisfied for some $T_c>0$ only if  $\alpha$ exceeds the threshold $\alpha_c=\Delta/2$. For  $\alpha<\alpha_c$ the system remains in a quantum paraelectric phase at $T=0$.

In the following we focus on the effect of the cavity on the paraelectric and quantum paraelectric phase. As discussed previously, the static susceptibility in the limit $N\to\infty$ does not depend on the light-matter interaction, and therefore the quantum paraelectric for $\alpha<\alpha_c$ cannot be turned into a true ferroelectric phase.  The  thermodynamics of a finite system, however, is affected by the coupling to the electromagnetic field, and one may anticipate a large effect on the susceptibility in particular close to the mean-field phase transition. To investigate this, we calculate the static susceptibility $\chi(0)$ for a finite system both using exact diagonalization and the analytic result for the leading $1/N$ correction to the exact result $\chi_{\infty}$ in the thermodynamic limit.

The effect of the cavity to leading order in $1/N$ could be obtained from the nonlinear susceptibilities of the interacting model outside the cavity, as derived in Sec.~\ref{sec:coll}. In the present case, also the interacting model outside the cavity is systematically solvable by means of an $1/N$ expansion. It is therefore more convenient to formulate the collective theory analogous to Sec.~\ref{sec:coll}, but directly for an all-to-all interaction which combines the interaction $\alpha$ and the light-induced interaction \eqref{vinddd} (see App.~\ref{app:all2all}). Within this approach, the $1/N$ correction can be evaluated in terms of a Hartree diagram for the self-energy analogous to Eq.~\eqref{pistatmaintext} (see App.~\ref{equ:int_hartree} for details). The propagator $W_0$ is then a mean-field propagator and diverges at the transition, so that the expansion works only in the quantum paraelectric phase and for $T>T_c$, which however is sufficient for our analysis.

\begin{figure}[h]
	\centerline{\includegraphics[width=0.5\textwidth]{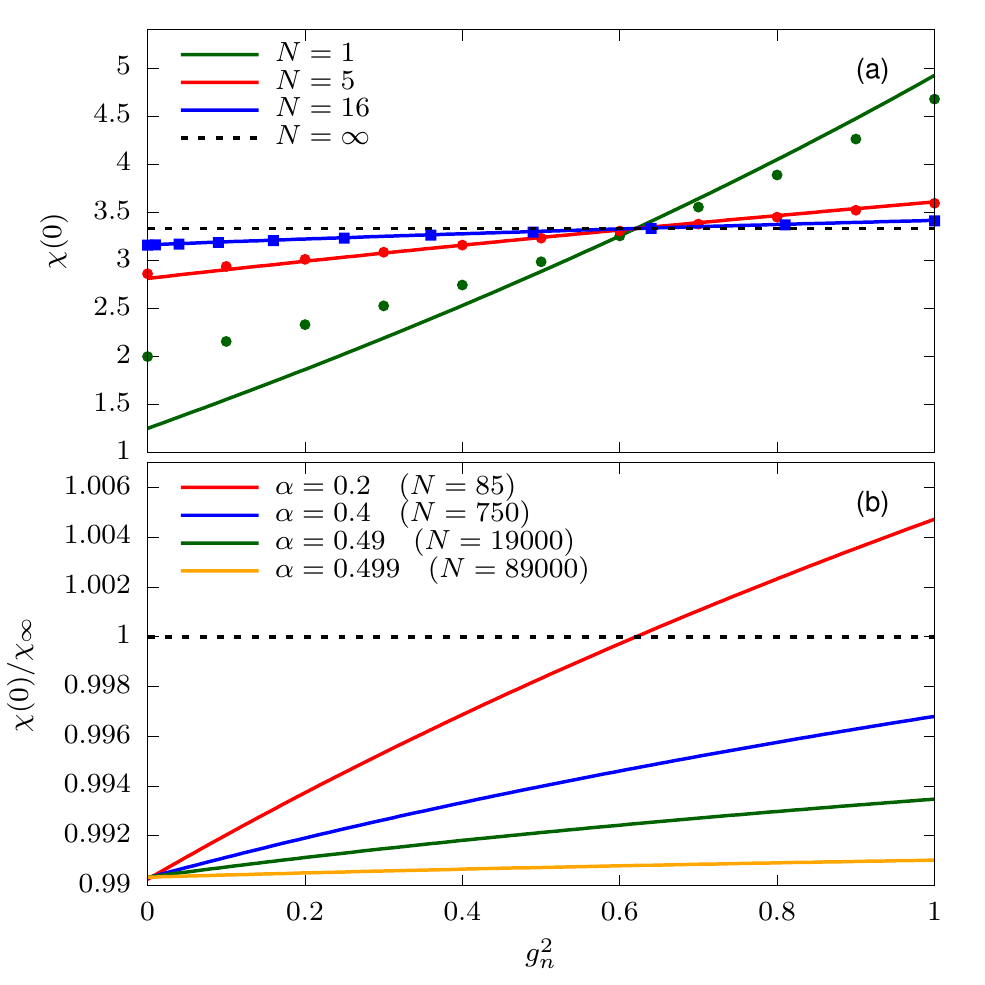}}
	\caption{
	(a) Static low-temperature  susceptibility of the interacting model as a function of the light-matter interaction strength $g_n^2$ in the quantum paraelectric regime ($\alpha=0.2$, $\Delta=1$, $\Omega=1$). The horizontal black dashed line indicates the mean-field result ($N\rightarrow\infty$). Dots correspond to numerical results obtained from exact diagonalization (ED) for a system size of $N=1$ (green) and $N=5$ (red). The data for $N=16$ represented by the blue squares have been obtained using the Lanczos method. The solid lines represent the analytic curve resulting from a direct evaluation of the Hartree diagram of the self-energy. (b) Low-temperature solution of the ratio $\chi(0)/\chi_{\infty}$ as a function of $g_n^2$ for various values of $\alpha$. The results have been calculated using the Hartree approximation for the self-energy. For each curve, the number of emitters has been chosen such that the relative deviation of $\chi(0)$ from the mean-field result is around $1\%$. 
}
	\label{fig:ata_lm_coupling_alpha0.2}
\end{figure}

\subsection{Quantum paraelectric phase}

\subsubsection*{Results}

Figure~\ref{fig:ata_temp_alpha0.2} shows the inverse static susceptibility (panel (a)) and the corresponding self-energy (panel (b)) for a system of five emitters and $\alpha=0.2$ at $g_n^2=0$ and $g_n^2=1$.  Here and in the following the unit of energy is set to the level splitting $\Delta$, and the cavity frequency is given by $\Omega=\Delta$, unless otherwise stated. The result $\chi_{\infty}^{-1}$ for the thermodynamic limit is indicated by the black dashed line, the solid curves represent the analytic $1/N$ result obtained from the  evaluation of the Hartree diagram, and the dots have been calculated using exact diagonalization. It is clearly visible that the result for the finite system depends on the light-matter interaction strength.  Without coupling to the cavity, finite size fluctuations decrease the susceptibility (increase $\chi^{-1}$) with respect to the $N=\infty$ result, while the cavity in turn increases the susceptibility. For sufficiently strong collective coupling $g_n^2$, the susceptibility can even be enhanced with respect to $\chi_{\infty}$.  In spite of the small number of emitters,  the analytical and exact diagonalization results are in remarkably good agreement. The inset in panel (a), which displays the $N$-dependence of the deviation $|\chi(0)-\chi_{\infty}|$ at $T=0.01$ (vertical gray line in panel (a)), confirms the $1/N$ convergence to the exact mean-field limit. 

The effect of the light-matter interaction becomes more evident in Fig.~\ref{fig:ata_lm_coupling_alpha0.2}(a), where the static susceptibility is plotted as a function of $g_n^2$ for $N=1$, $N=5$ and $N=8$, at fixed temperature $T=0.01$ and $\alpha=0.2$. The $N=\infty$ result does not depend on the light-matter coupling strength (dashed black line).  For finite $N$ and sufficiently large $g_n$, the cavity leads to an enhancement of the susceptibility even beyond the mean-field result, but  for fixed $g_n$ the quantitative correction to $\chi_\infty$ decreases with increasing $N$. It is now interesting to analyze the cavity effect when the system approaches the critical point $\alpha_c=0.5$, where $\chi_\infty$ diverges. In  Fig.~\ref{fig:ata_lm_coupling_alpha0.2}(b), we show the $g_n^2$-dependence of the relative  change $\chi(0)/\chi_\infty$ with respect to the thermodynamic limit for $\alpha=0.2$, $0.4$, $0.49$, $0.499$; the corresponding absolute values $\chi_\infty$ are 3.33,  10.0, 100.0 and 1000.0. In the critical regime $\alpha\lesssim\alpha_c$, increasingly large $N$ are needed for the leading $1/N$ correction to become accurate; for each $\alpha$, the value of $N$ is  therefore chosen such that the relative $1/N$ correction $|\frac{\chi_{\infty}-\chi(0)}{\chi_{\infty}}|$ at $g_n^2=0$ is already small ($\approx 1\%$). 

\begin{figure}
	\centerline{\includegraphics[width=0.5\textwidth]{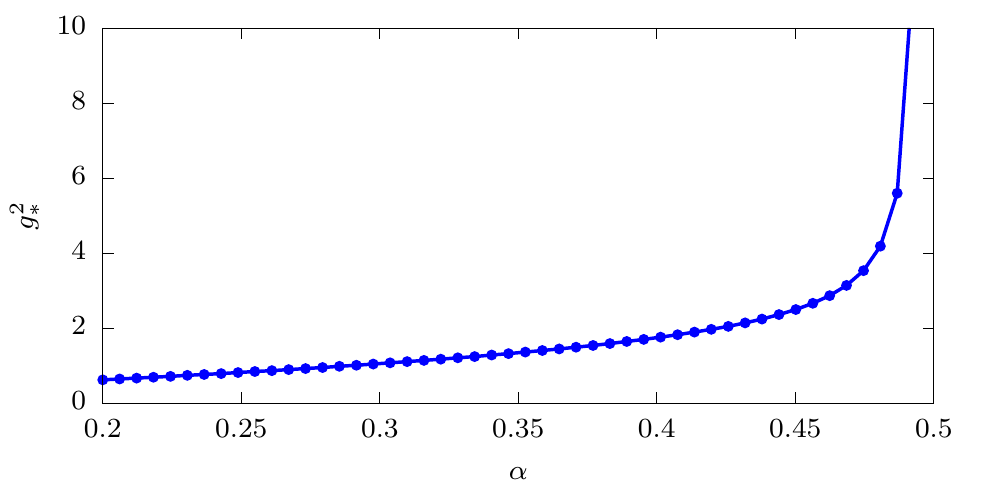}}
	\caption{ Value  $g_*^{2}$ of the coupling which is needed to balance the finite-size reduction of the static susceptibility (i.e., the crossing point of the static susceptibility $\chi(0)$ and the mean-field solution $\chi_{\infty}$) as a function of $\alpha$ at low temperature ($T=0.01$).}
	\label{fig:zero_of_pi}
\end{figure}

Again, the results prove that the light-matter interaction enhances the static susceptibility of the finite system. For given $g_n$, the effect is most pronounced for $\alpha=0.2$ and gets less and less significant as $\alpha$ is increased closer to criticality.  This statement can be rephrased as follows: At a given point $g_n\equiv g_*$, the cavity-induced enhancement of the susceptibility precisely balances the reduction due to finite-size fluctuations, $\chi(0)/\chi_{\infty}=1$ (see the point $g_*^2\approx 0.6$ for $\alpha=0.2$ in Fig.~\ref{fig:ata_lm_coupling_alpha0.2}(b)).  Note that $g_*$ is independent of $N$ within the $1/N$ theory. As a function of $\alpha$, the value $g_*$ becomes larger as $\alpha_c$ is approached, i.e., increasingly large light-matter couplings would be needed to balance the finite size fluctuations closer to the critical regime (Figure~\ref{fig:zero_of_pi}). This might be simply due to the fact that close to criticality finite-size fluctuations are more significant, such that one needs larger $g_n$ to compensate them. Moreover, the slowdown of the dynamics close to $\alpha_c$ may bring the system effectively closer to the high-cavity-frequency regime, which reduces the cavity effect as explained at the end of Section~\ref{subseccoll}.

\begin{figure}
	\centerline{\includegraphics[width=0.5\textwidth]{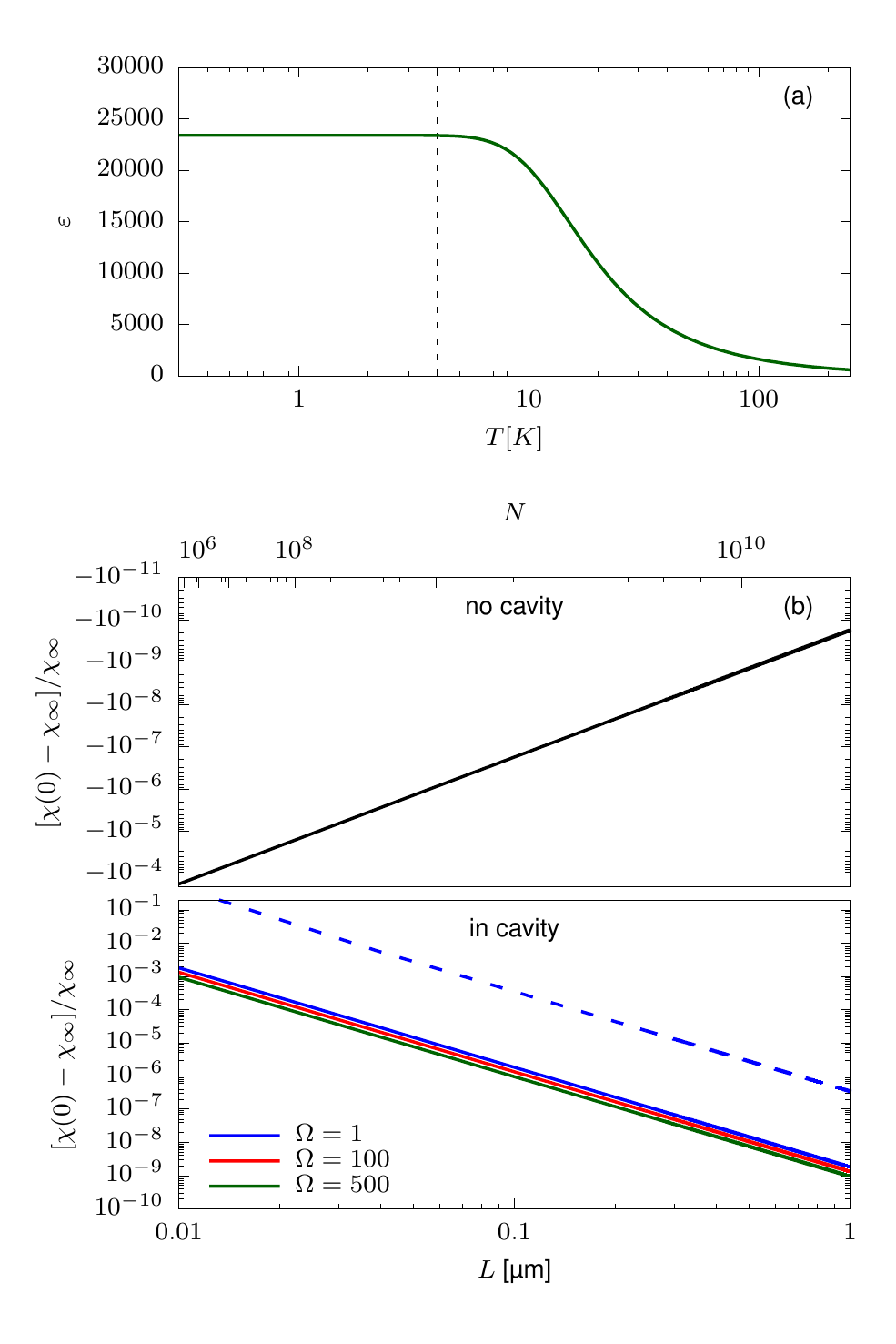}}
		\caption{
	(a) Dielectric constant as a function of $T$ for a system with $\alpha=0.328$ and $g_n^2=4024$. 	If the transition frequency is assumed to be equal to 5 THz ($\Delta=3.3$ meV), the resulting curve matches the one for STO. The black dashed line indicates the onset of the quantum paraelectric regime of STO at about 4 K. (b) Relative deviation of $\chi(0)$ from the mean-field limit for the parameters estimated for STO inside (lower panel) and outside (upper panel) the cavity for various cavity frequencies $\Omega$. The volume of the system is given by $L^3=Na^3$, where $a=3.9$ \AA~denotes the lattice constant of STO. The lower (upper) $x$-axis indicates the size of the system using the length scale $L$ (the number of emitters $N$). The blue dashed line shows the polaronic approximation for $\Omega=1$.}
	\label{fig:STO_eps}
\end{figure}

\subsubsection*{Comparison to STO}

 An important example of a real material that features a quantum paraelectric phase is the perovskite SrTiO\textsubscript{3} (STO) \cite{Mueller1979}. In this system, a metastable ferroelectric state has been induced by classical ultrashort laser pulses \cite{Nova2019,Li2019STO}, and STO has been proposed as a prime candidate for cavity-induced ferroelectricity \cite{Ashida2020,Latini2021}. We take this as a motivation to determine parameters of the minimal model \eqref{equ:Hmat_int} such that they reproduce the temperature-dependent bulk dielectric constant $\varepsilon(T)$ of STO. Within the present model,  $\varepsilon(T)$ is given by
\begin{equation}
\label{esfit}
\varepsilon(T)=1+g_n^2\chi_{\infty}(i\nu_m=0,T)
\end{equation} 
 (see App.~\ref{sec:eps} for more details). Note that this expression describes the material which is  not coupled to the cavity; the parameter $g_n^2$ appears because the same dipolar moments which determine the light-matter coupling also are responsible for the dielectric response. A fit of Eq.~\eqref{esfit} and \eqref{equ:chi_infty}  to the dielectric constant of STO taken from Ref.~\cite{Mueller1979} yields the parameters $\Delta=5$ THz, $\alpha=0.328\Delta$ and $g_n^2=4024\Delta$.  This three-parameter fit is analogous to the mean-field fit used in Refs.~\onlinecite{Barrett1952} and \onlinecite{Mueller1979}, although in the latter case the expression is obtained for a different atomic model (an anharmonic oscillator) within a mean-field approximation. The corresponding curve for $\varepsilon(T)$ is shown in Fig.~\ref{fig:STO_eps}(a).

We now use these parameters to calculate the relative deviation of the static susceptibility at low temperature from the thermodynamic limit for a finite system of volume $L^3=Na^3$, with the lattice constant $a=3.9$\AA{} of STO \cite{Shin2021}. The result is shown in Fig~\ref{fig:STO_eps}(b), where the lower and upper horizontal axis indicate the length $L$ in $\mu$m and the corresponding value of $N$, respectively. In the upper panel, we analyze the system without coupling to the cavity (the susceptibility computed at $g_n=0$), measuring the finite-size reduction on the static susceptibility. The lower panel displays the corresponding results with light-matter interaction ($g_n^2=4024$), which leads to an enhancement of the susceptibility. 
Note that the cavity effect is considerably larger than the finite size reduction at $g_n=0$. Both with and without the cavity, the correction to the bulk susceptibility decays like $1/N$ within the leading order theory. While an evaluation of higher order diagrams is beyond the scope of this article, one can assume that the leading order correction is accurate as long as it gives itself a small correction to $\chi_\infty$. For the present parameters, this holds down to very small cavities of $10$nm. Hence, within this mean-field model of the quantum paraelectric  the cavity effect on the static response becomes experimentally observable only in the limit of very small cavities. 

\subsubsection*{Polaronic interpretation}

A possible mechanism for the cavity enhancement of the susceptibility is based on dynamical localization \cite{Sentef2020}: The interaction with the cavity mode transforms the  polarizable degrees of freedom into light-matter polarons, which can effectively reduce the tunneling $\Delta$ between the classical configurations and therefore enhance the ordering tendency. It is illuminating  to analyze how this effect becomes quantitatively manifest in the present model. For this purpose, we perform an exact unitary transformation $\hat W$ to a Hamiltonian $\tilde H = \hat W H \hat W^\dagger$ in which the light-matter coupling enters via a renormalization of the tunneling term instead of the dipolar coupling. Using the standard Lang-Firsov polaron transformation
\begin{align}
\hat W= e^{ \frac{g_1}{\sqrt{2\Omega}} \hat P (a^\dagger -a)}
\end{align} 
with the total polarization $\hat P =\sum_{r} \hat \sigma_{x,r}$, the photon operators are shifted as $\hat W a \hat W^\dagger = a - g_1/\sqrt{2\Omega} \hat P$, and the light-matter Hamiltonian \eqref{square} is transformed to a decoupled form $\Omega a^\dagger a$ \cite{Bernardis2018}. In the new basis, the light-matter coupling instead arises from the transformed Hamiltonian $\hat W H_{\rm mat} \hat W^\dagger$. In total, we have
\begin{align}
\label{hpolaron}
\tilde H = \Omega a^\dagger a -\frac{\alpha}{2N}\sum_{r,r'} \hat{\sigma}_{x,r}\hat{\sigma}_{x,r'} + \sum_{r} \tilde h_{r}
\end{align}
where 
\begin{align}
\label{hren}
\tilde h_{r}
&=
\frac{\tilde \Delta}{2}
e^{\frac{g_1}{\sqrt{2\Omega}} \hat\sigma_{x,r} (a^\dagger -a)}\hat\sigma_{z,r} e^{-\frac{g_1}{\sqrt{2\Omega}}\hat\sigma_{x,r} (a^\dagger -a)}
\end{align}
is a single atom Hamiltonian where the light-matter interaction enters via photon absorption and emission in the tunneling.  The polaronic picture is then obtained by projecting the Hamiltonian \eqref{hren} to the zero photon sector (tunneling renormalization by virtual photons), which gives 
$\tilde h_{r} \approx\frac{\tilde \Delta}{2}\sigma_{z,r}$, with a renormalized tunneling
\begin{align}
\tilde \Delta  = \Delta e^{-g_1^2/2\Omega} =  \Delta e^{-\frac{1}{N}\frac{g_n^2}{2\Omega}}.
\end{align}
We can then evaluate the ``polaronic'' susceptibility $\chi_{\rm pol}$ for a system without cavity but with a renormalized tunneling $\tilde \Delta$. Also in this simple polaronic approach the cavity effect decreases like $1/N$, because the renormalization of $\Delta$ is controlled by the single-particle coupling $g_1^2$. Quantitatively, however, $\chi_{\rm pol}$ considerably overestimates the cavity-induced enhancement of $\chi$. This can be expected as on the one hand the projection to the polaronic Hamiltonian is valid only for $\Omega\gg \Delta$, and on the other hand the projection to the zero photon Hamiltonian at the level of a single atom neglects the light-induced interaction between the polarons. This emphasizes that the effect of cavity-induced localization on the phase transition has to be understood at the level of the collective theory \cite{Latini2021} and not on the single atom level.

\section{Conclusion}
\label{Sec:conclusion}

In conclusion, we have derived and analyzed a theory for the collective light-matter response in a many-particle system coupled to a single electromagnetic mode. Coupling to the single mode gives rise to a retarded all-to-all interaction in the solid. The all-to-all nature of this interaction allows for a controlled $1/N$ expansion of the effective theory for the collective modes. Nonlinearities in this theory, which are essential for any effect of the cavity on the static properties of the material  \cite{Latini2021,Ashida2020}, are given in terms of the nonlinear response functions of the material without coupling to the cavity. This allows to understand the properties of the material in the cavity in terms the uncoupled response, which can be measured or calculated by characterizing the system outside the cavity. The approach aims at the regime of strong collective coupling and weak single-particle coupling, complementary to numerical approaches for the regime of strong single-particle coupling \cite{Ashida2021PRL, Kim2021}. 

We have used the approach to discuss the renormalization of the ferroelectric response of a quantum paraelectric material, such as STO, in a single mode cavity. One finds a cavity-induced increase of the susceptibility, but only for relatively small clusters of the material in very small cavities where the mode volume is strongly compressed compared to the free space wavelength. Clearly, the simple mean-field model for the quantum paraelectric transition does not claim quantitative accuracy for any real material. In contrast, the general theory presented in our work explains how the response of the material in the cavity depends on the nonlinear response outside the cavity. This implies that a collective theory which is intended to predict the cavity response  beyond a trivial macroscopic description should in particular reproduce also the nonlinear response of the material outside the cavity, and not only  linear response quantities such as the dielectric constant. In a quantum paraelectric material, the collective tunnelling of large domains \cite{Fauque2022} will lead to a strong nonlinear response which is not quantitatively captured  within a mean-field description. It is therefore an interesting future research direction to extract the nonlinear response functions needed to compute the cavity response within our approach from a first principles theory for the quantum paraelectric phase \cite{Shin2021, Latini2021} that captures collective tunnelling.

The derivation of the collective theory relies solely on a linear coupling between light and matter, so it should generally  apply to order parameters such as incommensurate charge density wave or exciton condensates, which couple linearly to light. Moreover, the collective theory should be useful to compute the dynamic nonlinear response and higher order photon correlation functions in the cavity (extending on the description of linear response of strongly coupled light-matter hybrids \onlinecite{Flick2019}), and to interpret nonadiabatic QED experiments \cite{Halbhuber2020}.


\acknowledgements
We acknowledge useful discussions with C. Sch\"afer, S. Latini, A. Rubio, and D. Fausti. K.L and M. E. were funded by the ERC Starting Grant No. 716648, and by the Deutsche Forschungsgemeinschaft (DFG, German Research Foundation) – Project-ID 429529648 – TRR 306 QuCoLiMa (``Quantum Cooperativity of Light and Matter''). J.L. acknowledges the funding from the European Union’s Horizon 2020 research and innovation program under the Marie Skłodowska-Curie Grant No. 884104 and from SNSF Grant No. 200021-196966.

%

\appendix
\onecolumngrid

\section{Derivation of the collective theory}

\subsection{Imaginary time action}

As described in the main text, we consider a Hamiltonian
\begin{align}
\label{Hgeneral}
\hat{H}&=H_{\rm mat} +\hat{H}_{Ed} + \hat{H}_{dd}+\Omega a^\dagger a,
\,\,\,
\hat{H}_{Ed}= \sqrt{\Omega} g_1\hat X\hat P,\,\,\,\,
\hat{H}_{dd}=\frac{g_1^2}{2}\hat P^2,
\end{align}
where $H_{\rm mat}$ is an arbitrary Hamiltonian for a piece of matter consisting of $N$ atoms, $\hat{p}_r$ are dimensionless dipolar operators for atom $r$ and $\hat P=\sum_{r}\hat{p}_{r}$, $\hat X=(a^\dagger+a)/\sqrt{2}$, and $g_1 = \sqrt{d^2e^2/\epsilon V}$ is the single-particle coupling. For a field theoretical treatment of the problem described in the main text, we employ the imaginary-time path-integral formalism.  In the path-integral representation, the problem is defined in terms of variables $\eta(\tau)$ (such as coherent states, see below) and an action $S[\eta]$, such that the partition function is $Z=\int\mathcal{D}[\eta] e^{-S[\eta]}$, and expectation values are defined with respect to the action as $\langle X[\eta] \rangle_S = \frac{1}{Z} \int\mathcal{D}[\eta] e^{-S[\eta]} X[\eta]$. Multi-point expectation values  correspond to imaginary-time ordered correlation functions, 
\begin{align}
\langle X_1(\tau_1) X_1(\tau_2) \cdots \rangle_S
=
\langle T_\tau \hat X_1(\tau_1) \hat X_1(\tau_2) \cdots \rangle_H.
\end{align}
Here the right-hand side is the  operator-based expression for the time-ordered correlation function, with the thermal expectation value $\langle \cdots \rangle_H = \frac{1}{Z}\text{tr}[e^{-\beta \hat H} \cdots ]$, the time-ordering operator $T_\tau$, and the time-dependence of the operators in the Heisenberg picture, $\hat X (\tau) = e^{\tau\hat H} \hat X e^{-\tau\hat H}$. Moreover, two-time correlation functions $C(\tau_1,\tau_2) = \langle X_1(\tau_1) X_1(\tau_2) \rangle_S$ depend on time difference only, $C(\tau_1,\tau_2) = C(\tau_1-\tau_2)$ with the periodic function $C(\tau+\beta) = C(\tau)$ (we only need bosonic functions in the following). We therefore use the Matsubara frequency representation with frequencies $i\nu_m=2\pi n / \beta$,
\begin{align}
\label{cinun}
C(i\nu_m)&=\int_0^\beta d\tau\, C(\tau) e^{i\nu_m\tau},\,\,\,\,\,\,\,\,C(\tau) = \frac{1}{\beta}\sum_{m} e^{-i\nu_m\tau} C(i\nu_m).
\end{align}

The photonic variables are introduced as coherent state complex variables $a(\tau)$ and $\bar a(\tau)$ (complex conjugate). The free action, corresponding to the Hamiltonian $\hat H_\text{field}=\Omega \hat a^\dagger \hat a$ is given by
\begin{equation}
	S_\text{field}[\bar a,a]=\int_{0}^{\beta}d\tau  \,\bar a(\tau)(\partial_\tau+\Omega)a(\tau).
\end{equation}
The free photon propagator is given by
\begin{align}
\label{d0}
D_0(\tau) = -\langle a(\tau) \bar a(0)\rangle_{S_\text{field}} = -\frac{e^{-\tau\Omega}}{1-e^{-\beta\Omega}},
\end{align}
which is obtained either by the evaluation of the Gaussian path integral or by a straightforward operator representation. Fourier transform \eqref{cinun} gives 
\begin{align}
\label{d0w}
D_0(i\nu_m) = \frac{1}{i\nu_m-\Omega}.
\end{align}

The matter is described in terms of variables $c$, which do not have to be further specified here; they can be Grassmann variables if one chooses a Fermion representation of the two-level system, or a real  variable if the local degree of freedom is the displacement of some ion within the unit cell. We denote the action of the matter without light as $S_{\rm mat}[c]$. The remaining terms in the light-matter interaction  are
\begin{align}
S_{dd} + S_{Ed}&= \frac{g_1^2}{2}   \int_0^\beta  d\tau P(\tau)^2 + g_1 \sqrt{\frac{\Omega}{2}}\int_0^\beta d\tau\,  \big(a(\tau) + \bar a(\tau)\big)P(\tau),
\end{align}
with the collective dipole moment $P(\tau) =\sum_r p_{r}(\tau)$.
Furthermore, we add a source term
\begin{align}
S_{\eta}
&=
\int_0^\beta d\tau\, \big(a(\tau) \bar \eta(\tau)+ \bar  a(\tau)\eta(\tau)\big)
\end{align}
to later determine the photonic observables. 

\subsection{Light-induced interaction} 
\label{App:lightinduced}

The action for the isolated matter can be obtained by integrating out the photon variables. Writing the terms linear in the photonic fields as
\begin{align}
S_{Ed}+S_\eta = \int_0^\beta d\tau \big[a(\tau)\bar m(\tau) + \bar a(\tau) m(\tau)\big],
\end{align}
with $m(\tau)= \eta(\tau)+\sqrt{\Omega/2}\,g_1P(\tau)$, a Gaussian integral gives
\begin{align}
\label{actionmm}
&\frac{1}{Z_{field}}\int\mathcal{D}[\bar a,a]e^{-(S_\text{field} +S_{Ed}+ S_\eta)} =e^{-S_\text{eff}[c,\eta,\bar\eta]},
\\
&S_\text{eff}[c,\eta,\bar\eta]
=\int_{0}^\beta d\tau\int_{0}^\beta d\tau' \bar m(\tau)D_0(\tau-\tau')m(\tau'),
\end{align}
with the free photon propagator \eqref{d0}. The source free term of this effective action therefore defines an induced interaction $S_{\rm ind}=\tfrac12g_1^2\Omega \int_{0}^\beta\!\! d\tau \int_{0}^\beta \!\!d\tau' P(\tau)  D_0(\tau-\tau')P(\tau')$. It is convenient to write this term in symmetrized form, and combine it with the interaction $S_{dd}$, which gives an induced interaction
\begin{align}
\label{sindgehe}
S_{\rm ind}=&\frac{1}{2}\int_{0}^\beta\!\! d\tau \int_{0}^\beta \!\!d\tau'  P(\tau) 
\frac{1}{N}V_\text{ind}(\tau-\tau')
P(\tau'),
\end{align}
with 
\begin{align}
\label{vinducedjs}
V_{\rm ind}(\tau)&= \frac{g_n^2\Omega}{2} \big[D_0(\tau)+ D_0(\beta-\tau)\big] +g_n^2 \delta(\tau),
\,\,\,\,\,
V_{\rm ind}(i\nu_m)= g_n^2\frac{\nu_m^2}{\nu_m^2+\Omega^2}.
\end{align}
For later convenience, we have included the factor $N$ in the induced interaction (making it proportional to the collective coupling $g_n^2$ instead of the single particle coupling $g_1^2$), and in turn induced the factor $1/N$ in the integral \eqref{sindgehe}. Overall, the source-free matter-only action is therefore given by
\begin{align}
\label{effectiveca}
S_\text{eff}[c]&=S_{\rm mat} + S_{\rm ind}[c].
\end{align}
(To simplify the notation, we assume the sources $\eta$ to be zero in $S_\text{eff}$ unless they are shown explicitly.) We see that the light-induced interaction is zero at zero frequency, which will become important later.

\subsection{Exact relations for the photon observables}
\label{app:exactrelations01}

To relate the matter-only theory to the exact interacting photonic observables, we take derivatives with respect to the source fields. Using 
\begin{align}
Z[\eta,\bar\eta]
&
= \int \mathcal{D}[\bar a,a]\int\mathcal{D}[c] e^{-(S_{\rm mat} + S_{dd}+S_\eta + S_{ed}+S_\text{field})}= \int\mathcal{D}[c] e^{-(S_\text{eff}[c,\eta,\bar\eta] +S_{\rm mat}[c])},
\end{align}
and taking derivatives with respect to the sources on the left hand side gives 
\begin{align}
\langle a(\tau) \rangle_S
&=
-\frac{\delta}{\delta\bar \eta(\tau)}
\log Z[\eta,\bar \eta]\Big|_{\eta=\bar \eta=0}.
\end{align}
When we perform the derivatives using the action \eqref{actionmm}, we get
\begin{align}
\langle a(\tau) \rangle_S
&=
\int_{0}^\beta d\tau' D_{0}(\tau-\tau') g_1\sqrt{\frac{\Omega}{2}} \langle P(\tau')\rangle_S.
\end{align}
Assuming that the expectation values are static, 
\begin{align}
\langle a \rangle_S
&=
D_{0}(i\nu_0) g_1\sqrt{\frac{\Omega}{2}} \langle P\rangle_S
=
-\frac{g_1}{\sqrt{2\Omega}}
\langle P\rangle_S.
\end{align}
Similar we can obtain the connected photon propagator $D(\tau) = -\langle a(\tau) \bar a(\tau')\rangle^\text{con}_{S}$ (the superscript indicates that this is the connected correlation function), by taking derivatives with respect to the source fields $\eta$,
\begin{align}
D(\tau-\tau') 
&=
-\frac{\delta}{\delta\bar \eta(\tau)}
\frac{\delta}{\delta\eta(\tau')}
\log Z[\eta,\bar \eta]\Big|_{\eta=\bar \eta=0}.
\end{align}
When we take the derivatives using the action \eqref{actionmm}, we get
\begin{align}
D(i\nu_m)=
D_0(i\nu_m)
-
D_0(i\nu_m)
\Big[
 \frac{\Omega g_n^2}{2}
\chi(i\nu_m)
\Big]
D_0(i\nu_m),
\end{align}
where $\frac{1}{N}\chi(\tau)=\langle P(\tau) P(0)\rangle_S^\text{con}$ is the interacting susceptibility. Note that the term $\frac{1}{N}\sum_{r,r'} \chi_{r,r'}(i\nu_m)$ is $\mathcal{O}(1)$ for large $N$, so that one can see that the modification of the photon propagator is controlled by the collective coupling $g_n^2$ instead of the single particle coupling $g_1^2$. 

\subsection{Hubbard Stratonovich representation}
\label{App:HubStrato}

We now attempt to derive a model for collective degrees of freedom only. We can decouple the induced interaction \eqref{sindgehe} by a single real Hubbard Stratonovich field. Since the matrix $V_{\rm ind}$ is positive definite, we can use the identity
\begin{align} 
\label{actionphiwgw}
e^{-S_{\rm ind}}=
\frac{1}{Z_V}
\int \mathcal{D}[\varphi] e^{
-\frac{1}{2}\int_0^\beta d\tau \int_0^\beta d\tau' \varphi(\tau)V_{\rm ind}^{-1}(\tau-\tau')\varphi(\tau')
-i\int_0^\beta d\tau \sum_{r} \frac{\varphi(\tau)}{\sqrt{N}} p_{r}(\tau)},
\end{align}
to represent the induced interaction \eqref{sindgehe}, with an irrelevant constant $Z_V$.  Before proceeding with a treatment of this action, let us summarize the exact relation between the interacting $\varphi$ propagator, which we define as
\begin{align}
W(\tau) = \langle \varphi(\tau) \varphi(0)\rangle^\text{con}_S,
\end{align}
and the matter susceptibility. With the generating function
\begin{align}
\label{generatinh}
Z[\xi] = \log \int\mathcal{D}[c]\int\mathcal{D}[\varphi] e^{-S} e^{\frac{i}{\sqrt{N}}\int_0^\beta d\tau \xi(\tau) \sum_{r}p_{r}},
\end{align}
we have
\begin{align}
\label{derivbbbY}
\chi(\tau-\tau') 
\equiv
\frac{1}{N}\sum_{r,r'} \langle p_{r}(\tau) p_{r'}(\tau')\rangle^\text{con}_S
&=
-\frac{\delta}{\delta\xi(\tau)}
\frac{\delta}{\delta\xi(\tau')}
\log Z[\xi]\Big|_{\xi=0}.
\end{align}
At the same time, we can perform a shift of the integration variable $\varphi$ to $\tilde \varphi = \varphi-\xi$ in Eq.~\eqref{generatinh}, so that the generating function reads
\begin{align}
\label{generatinh01jj}
Z[\xi] = \log \int\mathcal{D}[c]\int\mathcal{D}[\tilde\varphi] 
e^{-\big(S_{\rm mat}
+ \frac{1}{2}\int_0^\beta d\tau \int_0^\beta d\tau' (\tilde\varphi(\tau)+\xi(\tau))V^{-1}_{\rm ind}(\tau-\tau')(\tilde\varphi(\tau')+\xi(\tau'))
+\frac{i}{\sqrt{N}}\int_0^\beta d\tau \tilde\varphi(\tau)P(\tau)\big)}. 
\end{align}
Taking the derivatives as in Eq.~\eqref{derivbbbY} now gives
\begin{align}
\label{derivbbbY01}
\frac{\delta}{\delta\xi(\tau)}
\frac{\delta}{\delta\xi(\tau')}
\log Z[\xi]\Big|_{\xi=0}
=
-V_{\rm ind}^{-1}(\tau-\tau')
+
\int_0^\beta d\tau_1
\int_0^\beta d\tau_2
V_{\rm ind}^{-1}(\tau-\tau_1)
W(\tau_1-\tau_2)
V_{\rm ind}^{-1}(\tau_2-\tau').
\end{align}
Hence, in frequency space, we have the relations,
\begin{align}
\label{vgehjeveee01}
\chi(i\nu_m)
&=
V_{\rm ind}(i\nu_m)^{-1}
-
V_{\rm ind}(i\nu_m)^{-1}
W(i\nu_m)
V_{\rm ind}(i\nu_m)^{-1},
\\
\label{vgehjeveeeww}
W(i\nu_m)
&=
V_{\rm ind}(i\nu_m)
-
V_{\rm ind}(i\nu_m)
\chi(i\nu_m)
V_{\rm ind}(i\nu_m).
\end{align}

Next, we can integrate out the matter in Eq.~\eqref{generatinh01jj}. In general, the function
\begin{align}
\mathcal{G}_{\rm mat}[y]
=
-
\frac{1}{N}\log
\int\mathcal{D}[c_r] e^{-\big(S_{\rm mat} + i\sum_{r}\int_0^\beta d\tau p_r(\tau)y(\tau)\big)} 
\end{align}
is identified as the generating function for connected 
collective correlation functions
\begin{align}
\frac{\delta}{\delta y(\tau_1)}
\cdots
\frac{\delta}{\delta y(\tau_n)}
\mathcal{G}_{\rm mat}[y]\Big|_{y=0}
=
-
\frac{(-i)^n}{N}
\sum_{r_1,...,r_n}
\langle
p_{r_1}(\tau_1)\cdots
p_{r_n}(\tau_n)
\rangle^\text{con}
\equiv
\chi^{(n)}_{\rm mat}(\tau_1,...,\tau_n).
\label{chicondef}
\end{align}
(Below we denote $\chi_{\rm mat}^{(2)}=\chi_{\rm mat}$.) We can therefore write the function formally in a Taylor series
\begin{align}
\mathcal{G}_{\rm mat}[y]
=
\sum_{n=2,4,...}
\frac{1}{n!}
 \int_0^\beta d\tau_1\cdots d\tau_n\,
y(\tau_1)
\cdots
y(\tau_n)
\,
\chi^{(n)}_{\rm mat}(\tau_1,...,\tau_n),
\label{gejdkde}
\end{align}
where we have omitted an irrelevant constant (zeroth order term). Moreover, for symmetry reasons, we expect the action $S_{\rm mat}$ to be invariant under an inversion (which reverses the sign of $p_r$), so that only even orders contribute. Hence, the action \eqref{generatinh01jj} after integrating out the matter gives
\begin{align} 
\label{actioesphiwgw}
&S= \frac{1}{2}\int_0^\beta d\tau \int_0^\beta d\tau' \varphi(\tau)V_{\rm ind}^{-1}(\tau-\tau')\varphi(\tau')
+N {\mathcal{G}_{\rm mat}}[\varphi/\sqrt{N}].
\end{align}
The quadratic term in $N {\mathcal{G}_{\rm mat}}[\varphi/\sqrt{N}]$, which is given by $\frac{1}{2}\int_0^\beta d\tau \int_0^\beta d\tau' \varphi(\tau)\chi_{\rm mat}(\tau-\tau')\varphi(\tau')$, can  be combined with the first term, so that the final action reads
\begin{align} 
\label{actioespddgw01}
S&= \frac{1}{2}\int_0^\beta d\tau \int_0^\beta d\tau' \varphi(\tau) W_0^{-1}(\tau-\tau')\varphi(\tau')
+
\frac{1}{N} S^{(4)} + \mathcal{O}(1/N^2),
\\
S^{(4)}
&=\frac{1}{4!}\int_{1234} \chi^{(4)}_{\rm mat}(1,2,3,4) \varphi_1\varphi_2\varphi_3\varphi_4,
\end{align}
where 
\begin{align}
W_0^{-1} = V_{\rm ind}^{-1}+\chi_{\rm mat} \,\,\,\,\Rightarrow\,\,\,\,
W_0=\frac{V_{\rm ind}}{1+V_{\rm ind}\chi_{\rm mat}}.
\label{W0}
\end{align}
In the examples below, we can confirm that all signs are correct, in the sense that the quadratic term $W_0$ is again positive, and the Gaussian integral is convergent. The action \eqref{actioespddgw01} is the central result Eq.~\eqref{collectiveaction} in the main text.

\subsection{Perturbation series}
\label{App:parturbation}

The exact exaction \eqref{actioespddgw01} can be taken as a starting point for a diagrammatic expansion for the interacting Hubbard Stratonovich propagator $W$. The noninteracting propagator, from the quadratic action, is given by Eq.~\eqref{W0}. For the corrections we define a self-energy,
\begin{align}
\label{wdyson}
W^{-1} = W_{0}^{-1}- \Pi.
\end{align}
Going back to \eqref{vgehjeveee01}, the matter susceptibility would be
\begin{align}
\chi = V_{\rm ind}^{-1} - V_{\rm ind}^{-1} \frac{1}{V_{\rm ind}^{-1}+\chi_{\rm mat}-\Pi}V_{\rm ind}^{-1} 
=
\frac{\chi_{\rm mat}-\Pi}{1+(\chi_{\rm mat}-\Pi)V_{\rm ind}}.
\end{align}
Since the static $V_{\rm ind}(i\nu_0)=0$, we have
\begin{align}
\chi(i\nu_0)=
\chi_{\rm mat}(i\nu_0)-\Pi(i\nu_0),
\end{align}
i.e., the static self-energy directly gives the correction to the static susceptibility. 

The structure of the diagrammatic expansion is discussed in the main text. In particular, we see that higher-order self-energy diagrams  (beyond the leading Hartree diagram which is discussed below) can be neglected in two relevant limits: (i) When $N$ is large and $g_n^2$ is comparable to atomic energy scales in matter  (the usual limit for many particles in the cavity), and (ii), when $N$ is small but the system is not in the ultrastrong single-particle coupling regime  ($g_1^2$ is small). The extreme case of (i) would be the thermodynamic limit. There the mode volume is increased at given particle density ($L\to\infty$, $N\to\infty$, $N/L^3$ fixed), so that $g_n^2$ is fixed while $N\to\infty$, and thus the $1/N$ corrections entirely vanish. Hence, it is correct to say that in this limit, the behavior of the matter in the cavity is described by the zeroth order collective theory $\Pi=0$, which implies
\begin{align}
\chi = 
\frac{\chi_{\rm mat}}{1+\chi_{\rm mat}V_{\rm ind}}.
\end{align}
This is precisely the mean field expression \eqref{chimf} quoted in the main text, and which is also obtained from a straightforward mean-field decoupling of the induced interaction term in the matter-only theory \eqref{effectiveca}.

\subsection{Hartree diagram}

\label{App:HArtree}

Standard diagrammatic rules for the Hartree diagram (first diagram in Fig.~\ref{figdiagrams} in the main text) give
\begin{align}
\label{picorrection}
\Pi(\tau_1,\tau_2) =
-\frac{1}{2N}\int_0^\beta d\tau_3 d\tau_4\,\chi^{(4)}_{\rm mat}(\tau_1,\tau_2,\tau_3,\tau_4) W_0(\tau_3,\tau_4).
\end{align}
A nice aspect about Eq.~\eqref{picorrection} is that it relates the effect of the cavity to properties of the material without coupling to the cavity. It is therefore convenient to rewrite the formula in terms of measurable real-frequency quantities. A simplification is possible in particular for the static ($i\nu_0$) quantities. Starting from \eqref{picorrection}, and using translational invariance in time, we can also write 
\begin{align}
\Pi_\text{stat} =\frac{1}{\beta}\int_0^\beta d\tau_1d\tau_2 \, \Pi(\tau_1,\tau_2)
=
-\frac{1}{2N\beta}\int_0^\beta d\tau_1d\tau_2d\tau_3d\tau_4\, \chi_{\rm mat}^{(4)}(\tau_1,\tau_2,\tau_2,\tau_3)W_{0}(\tau_3-\tau_4).
\label{egfehe}
\end{align}
With the definition of the connected correlation functions as derivatives \eqref{chicondef}, we note that that the $\tau$-integrated derivatives can be replaced by derivatives with respect to a static field, 
\begin{align}
\int_0^\beta d\tau_1 \frac{\delta }{\delta y(\tau_1)} 
\log
\int\mathcal{D}[c_r] e^{-\big(S_{\rm mat} + i\sum_{r}\int_0^\beta d\tau p_r(\tau)y(\tau)\big)}
=
i\frac{\partial}{\partial h} 
\log
\int\mathcal{D}[c_r] 
e^{-S_{\rm mat}(h)},
\end{align}
where $S_{\rm mat}(h)=S_{\rm mat} + h \sum_{r}\int_0^\beta d\tau p_r(\tau)$ is the action including the static field $h$.
This can be used to remove the $\tau_1$ and $\tau_2$ integrations in Eq.~\eqref{egfehe},
\begin{align}
\Pi_\text{stat} 
=
-\frac{1}{2N\beta}
\int_0^\beta d\tau_3d\tau_4\, W_{0}(\tau_3-\tau_4)\,
(i)^2\frac{\partial^2}{\partial h^2}  \frac{\delta^2}{\delta y(\tau_3) \delta y(\tau_4)} 
\Big(\frac{-1}{N}\Big)\log
\int\mathcal{D}[c_r] e^{-\big(S_{\rm mat}(h)+ i\sum_{r}\int_0^\beta d\tau p_r(\tau)y(\tau)\big)}
\Big|_{y=0,h=0}.
\end{align}
The second derivative gives the connected correlation function in the presence of the fields $h$,
\begin{align}
\frac{\delta^2}{\delta y(\tau_3) \delta y(\tau_4)} 
\Big(\frac{-1}{N}\Big)\log
\int\mathcal{D}[c_r] e^{-\big(S_{\rm mat}(h)+ i\sum_{r}\int_0^\beta d\tau p_r(\tau)y(\tau)\big)}
\Big|_{y=0}
=
\frac{1}{N}\sum_{r,r'} \langle p_{r}(\tau_3) p_{r'}(\tau_4)\rangle_{h}^\text{con}
\equiv
\chi_{h}(\tau_3-\tau_4),
\end{align}
where $\langle\cdots\rangle_{h}=\langle\cdots\rangle_{S_{\rm mat}(h)}$ is the expectation value in the presence of the field (but without cavity). Hence, we have
\begin{align}
\Pi_\text{stat} 
=
\frac{1}{2N\beta}
\frac{\partial^2}{\partial h^2} 
\int_0^\beta d\tau_3d\tau_4\, W_{0}(\tau_3-\tau_4)\chi_{h}(\tau_3-\tau_4)\Big|_{h=0}.
\end{align}
To get the final result, we again use translational invariance in time, and the symmetry $W(\tau)=W(\beta-\tau)$,
\begin{align}
\Pi_\text{stat} =
\frac{1}{2N}
\frac{\partial^2}{\partial h^2} 
\int_0^\beta d\tau \,W_{0}(\beta-\tau)\chi_{h}(\tau)\Big|_{h=0}.
\label{pistat}
\end{align}
Given that one computes (or measures) the matter susceptibility in a static field outside the cavity, this allows to compute the modification of the susceptibility in the cavity. 

For a further rewriting we can use a general spectral decomposition into real bosonic correlation functions. In general, we attempt a decomposition (with real $G_{0}$, $G_{\infty}$)
\begin{align}
&G(i\nu_m) = \sum_{l} \frac{A_l}{i\nu_m-\Omega_l} + \delta_{n,0}G_{s} + G_{\infty} \equiv \tilde G(i\nu_m)+\delta_{n,0}G_\text{stat} + G_{\infty},
\label{gbosoninun}
\end{align}
where $G_s$ and $G_\infty$ contain possible $\delta$-function contributions in frequency at $\omega=0$ and in in time, and  $\tilde G(i\nu_m)$ is the regular part. The function $G_{\infty}$ is simply the high-frequency limit of $G$. A Fourier transform relates \eqref{gbosoninun} to 
\begin{align}
&G(\tau) = \sum_{l} A_l b(-\Omega_l)e^{-\Omega_l\tau} + T G_{s} + \delta(\tau) G_{\infty}  \equiv \tilde G(\tau)+ T G_{0s} + \delta(\tau) G_{\infty} 
\\
&\Leftrightarrow\,\,\,
\tilde G(\tau)=\int d\omega A(\omega) b(-\omega)e^{-\omega\tau}
\text{~with~}
A(\omega)=-\frac{1}{\pi}\text{Im}\tilde G(\omega+i0)=-\frac{1}{\pi}\tilde G''(\omega).
\end{align}
For the real bosonic correlation functions, we have pairs of poles at positive and negative energies with opposite sign.

In general, for $\chi_h$ we do not expect an instantaneous response, such that $\chi_{h,\infty}=0$,  because matter does not respond at frequencies much higher than its intrinsic energies. If $\chi_{\rm mat}$ has zero high-frequency limit, then by Eq.~\eqref{W0}, the high-frequency limit of $W_0$ is the same as for $V_{\rm ind}$,
\begin{align}
W_{0,\infty} = g_n^2.
\end{align}
The regular part of $W_0$ is just the difference
\begin{align}
\tilde W_0(i\nu_m) 
&= 
\frac{V_{\rm ind}(i\nu_m)}{1+V_{\rm ind}(i\nu_m)\chi_{\rm mat}(i\nu_m)} - g_n^2
=
- g_n^2
\frac{
\Omega^2+ g_n^2 \nu_m^2\chi_{\rm mat}(i\nu_m)
}{\nu_m^2+\Omega^2+ g_n^2 \nu_m^2\chi_{\rm mat}(i\nu_m)},
\label{egdjerjre}
\end{align}
which obviously behaves like $\sim 1/\nu_m^2$ for large $n$, since $\chi_{\rm mat}(i\nu_m)\sim 1/\nu_m^2$. With this representation of  Eq.~\eqref{pistat}, we can split
\begin{align}
\Pi_\text{stat} &= \tilde \Pi_\text{stat} + \Pi_\text{stat}^\infty,
\\
\label{pistattau}
\tilde \Pi_\text{stat}
&=
\frac{1}{2N}
\frac{\partial^2}{\partial h^2} 
\int_0^\beta d\tau \,\tilde W_{0}(\beta-\tau)\chi_{h}(\tau)\Big|_{h=0},
\\
\Pi_\text{stat}^\infty
&=
\frac{g_n^2}{2N}
\frac{\partial^2}{\partial h^2} \,\chi_{h}(\tau=0)\Big|_{h=0}.
\label{psisttainfty}
\end{align}
This equation should be suitable for a numerical evaluation: Assuming that $\chi_h$ can be computed, e.g., by exact numerics, DMFT, etc., one can then get $\chi_{h}(i\nu_m)$, and from that $\tilde W_0(i\nu_m)$ via Eq.~\eqref{egdjerjre}. The convolution can then, e.g., be evaluated in frequency space as
\begin{align}
\label{pistatinun}
\tilde \Pi_\text{stat}
&=
\frac{1}{2N}
\frac{\partial^2}{\partial h^2} 
T\sum_{m}  \tilde W_{0}(i\nu_m)\chi_{h}(i\nu_m)\Big|_{h=0}.
\end{align}

\subsection{Dicke model}
\label{App:Dike}

As a first problem, let us study $N$ independent two-level systems. This is the original Rabi problem, where the atoms interact only via the light. 
We first note that for the system of isolated atoms, all connected correlation functions are the same as for the ensemble (because the collective functions $\chi_{\rm mat}$ in Eq.~\eqref{chicondef} are defined {\em per atom}, i.e., with a factor $1/N$).  We therefore first compute the $h$-dependent correlation function from the atomic Hamiltonian
\begin{align}
H_{at}=\frac{\Delta}{2}\sigma_z + h \sigma_{x}.
\end{align}
Let us first calculate the susceptibility for $h=0$, 
\begin{align}
\chi_\text{at}(\tau) = \frac{1}{Z}\text{tr}\big(e^{-(\beta-\tau)H}\sigma_{x} e^{-\tau H}\sigma_{x}\big)
=
\frac{e^{(\beta-\tau) \Delta/2} e^{-\tau \Delta/2}+e^{-(\beta-\tau) \Delta/2} e^{\tau \Delta/2}
}{e^{\beta\Delta/2}+e^{-\beta\Delta/2}}
=
\frac{2\cosh((\beta-2\tau)\Delta/2)}{2\cosh(\beta\Delta/2)}.
\label{chattau}
\end{align}
Thus 
\begin{align}
\int_0^\beta d\tau e^{i\nu_m\tau}  \chi_{at}(\tau)=\tanh\Big(\frac{\beta\Delta}{2}\Big)
\Big[
\frac{1}{{i\nu_m+\Delta}}-\frac{1}{{i\nu_m-\Delta}}
\Big]
=
\tanh\Big(\frac{\beta\Delta}{2}\Big)
\frac{2\Delta}{\Delta^2+\nu_m^2}.
\end{align}

For $h\neq 0$, there are two eigenstates $|\pm\rangle$ with energy $E_{\pm}=\pm\sqrt{\Delta^2+4h^2}/2\equiv\pm E_h/2$. Relevant matrix elements are $|v|^2\equiv |\langle+|\sigma_x|-\rangle|^2=\Delta^2/E^2$ and $|u|^2\equiv |\langle+|\sigma_x|+\rangle|^2=|\langle-|\sigma_x|-\rangle|^2=1-|v|^2$. With this we have, using $\chi_h(\tau)=\langle \sigma_x(\tau)\sigma_x(0)\rangle_h-\langle \sigma_x\rangle_h^2$, and a similar calculation as for $h=0$,
\begin{align}
\chi_h(i\nu_m) 
&= |v|^2\tanh\Big(\frac{\beta E_h}{2}\Big) \frac{2E_h}{E_h^2+\nu_m^2}
+
\delta_{n,0}\beta (1-|v|^2)\Big[1-\tanh\Big(\frac{\beta E_h}{2}\Big)^2\Big]
\label{chihat}
\\
&=
 \frac{\Delta^2}{E_h}\tanh\Big(\frac{\beta E_h}{2}\Big) \frac{2}{E_h^2+\nu_m^2}
+
\delta_{n,0} \beta\frac{4h^2}{E_h^2}\Big[1-\tanh\Big(\frac{\beta E_h}{2}\Big)^2\Big],
\label{chihat007}
\end{align}
and
\begin{align}
\chi_h(\tau) 
&=
\frac{4h^2}{E_h^2}\Big[1-\tanh\Big(\frac{\beta E_h}{2}\Big)^2\Big]
+
 \frac{\Delta^2}{E_h^2} \frac{\cosh((\frac{\beta}{2}-\tau)E_h)}{\cosh(\beta E_h/2)}.
 \label{chitwhhee}
\end{align}

With this we can compute $\Pi_\text{stat}^\infty$ from Eq.~\eqref{psisttainfty}
\begin{align}
\Pi_\text{stat}^\infty
&=
\frac{g_n^2}{2N}
\frac{\partial^2}{\partial h^2} 
\Big[
1-\frac{4h^2}{E_h^2} \tanh^2\Big(\frac{\beta E_h}{2}\Big)\Big]_{h=0}
=
-\frac{4g_n^2}{N}
\frac{1}{\Delta^2} \tanh^2\Big(\frac{\beta\Delta}{2}\Big).
\label{psisttainftyat}
\end{align}
One can nicely see how this is a  correction to the free susceptibility $\chi_{\rm mat}(i\nu_0)$ which is perturbative in $g_1^2/\Delta$.

To compute the contribution $\tilde \Pi_\text{stat}$, we derive a spectral representation for $\tilde W_0$. For $W_0$ we have (using the for $\chi_{\rm mat}$ the expression Eq.~\eqref{chihat} at $h=0$),
\begin{align}
W_0
&=\frac{V_{\rm ind}}{1+V_{\rm ind}\chi_{\rm mat}} 
= \frac{g_n^2 \nu_m^2}{\nu_m^2+\Omega^2 +  g_n^2 \nu_m^2 \tanh(\beta\Delta/2)\frac{ 2\Delta}{\Delta^2+\nu_m^2}}
\\
&=g_n^2  \frac{\nu_m^2 (\Delta^2+\nu_m^2)}{(\nu_m^2+\Omega^2)(\Delta^2+\nu_m^2) + 2\nu_m^2 g_n^2 \Delta\tanh(\beta\Delta/2)}. 
\label{fwlehekek}
\end{align}  
We look for zeros of the dominator, which will correspond to the hybrid mode energies. These are given by the solutions of the quadratic equation for the variable $(\nu_m^2)$,
\begin{align}
(\nu_m^2+\Omega^2)(\Delta^2+\nu_m^2) + \nu_m^2 2 g_n^2 \Delta \tanh(\beta\Delta/2)=0.
\end{align}
By construction there are two negative solutions, denoted by $\nu_m^2=-x_{\sigma}^2$, with $\sigma=\pm$:
\begin{align}
x_{\pm}^2
=
\frac{1}{2}\Big[
(\Omega^2+\Delta^2+2 g_n^2 \Delta \tanh(\beta\Delta/2))\pm
\sqrt{
(\Omega^2+\Delta^2+2 g_n^2 \Delta \tanh(\beta\Delta/2))^2-4\Omega^2\Delta^2
}
\Big].
\end{align}
For $g_n=0$, these correspond to the unrenormalized eigenenergies $x_{+}=\Omega$ and $x_{-}=\Delta$ or vice versa.  
Hence the denominator in Eq.~\eqref{fwlehekek} is $(\nu_m^2+x_+^2)(\nu_m^2+x_-^2)$, and with a partial fraction decomposition 
\begin{align}
W_0
&=g_n^2  \frac{\nu_m^2 (\Delta^2+\nu_m^2)}{x_-^2-x_+^2}
\Big[\frac{1}{\nu_m^2+x_+^2} - \frac{1}{\nu_m^2+x_-^2}\Big]
\\
&=-\frac{g_n^2}{2} \frac{i\nu_m (\Delta^2+\nu_m^2)}{x_-^2-x_+^2}
\Big[
\frac{1}{x_+-i\nu_m} - \frac{1}{x_++i\nu_m}
-\frac{1}{x_--i\nu_m} + \frac{1}{x_++i\nu_m}\Big].
\end{align}  
One can indeed confirm that the high-frequency limit of this expression is $g_n^2$. Upon analytical continuation $i\nu_m\to\omega+i0$ (keeping $i0$ only where necessary), we get the imaginary part of the regular contribution $\tilde W_0$,
\begin{align}
\tilde W''_0(\omega) 
&=
-\frac{g_n^2}{2} \text{Im} \frac{\omega(\Delta^2-\omega^2)}{x_-^2-x_+^2}
\Big[
\frac{1}{x_+-(\omega+i0)} - \frac{1}{x_++(\omega+i0)}
-\frac{1}{x_--(\omega+i0)} + \frac{1}{x_++(\omega+i0)}\Big]
\\
&=
\pi \frac{g_n^2}{2}  \frac{\omega(\Delta^2-\omega^2)}{x_-^2-x_+^2}
\Big[
\delta(\omega-x_+)
+\delta(\omega+x_+)
-\delta(\omega-x_-)
-\delta(\omega+x_-)
\Big]
\\
&=
-\pi \frac{g_n^2}{2}
\sum_{\sigma\in\pm}
\frac{\sigma x_\sigma(\Delta^2-x_\sigma^2)}{x_-^2-x_+^2}
\big(\delta(\omega-x_\sigma) -\delta(\omega+x_\sigma) \big).
\end{align}
Hence we have the mode decomposition
\begin{align}
\tilde W''_0(\omega) 
&=
-\pi \frac{g_n^2}{2}
\sum_{\sigma\in\pm}
A_\sigma
\big(\delta(\omega-x_\sigma) -\delta(\omega+x_\sigma) \big),
\text{~~~with~~~}
A_\sigma
=
\frac{\sigma x_\sigma(\Delta^2-x_\sigma^2)}{x_-^2-x_+^2}.
\end{align}  
As a cross-check, we note that $A_\sigma>0$; $W_0$ thus described the coupling to the two hybrid modes, with respective coupling strength $A_\sigma$. For $g_n^2 \to 0$, the modes reduce to $\Omega$  and $\Delta$, with the coupling $A_\Delta=0$ and $A_{\Omega} =\Omega$, as it should be, because in this limit $\tilde W_0$ is the regular part of $V_{\rm ind}$. Moreover, we have the sum-rule 
\begin{align}
\label{sumrulea}
\sum_{\sigma} \frac{A_\sigma}{x_\sigma} =1,
\end{align}
which eventually implies that $\tilde W_0(i\nu_0) =  -g_n^2$, and thus $W_0(i\nu_0)=0$, as it should be.
In imaginary time, 
\begin{align}
\tilde W''_0(\tau) 
&=
-\frac{g_n^2}{2}
\sum_{\sigma\in\pm}
A_\sigma
\frac{\cosh((\frac{\beta}{2}-\tau)x_\sigma)}{\sinh(\beta x_\sigma/2)}.
\end{align}  
This can now be combined in Eq.~\eqref{pistattau} with Eq.~\eqref{chitwhhee},
\begin{align}
\tilde \Pi_\text{stat}
&=
-\frac{g_n^2}{4N}
\sum_{\sigma\in\pm}
A_\sigma
\frac{\partial^2}{\partial h^2} 
\int_0^\beta d\tau 
\frac{\cosh((\frac{\beta}{2}-\tau)x_\sigma)}{\sinh(\beta x_\sigma/2)}
\Big[\frac{4h^2}{E_h^2}\Big[1-\tanh\Big(\frac{\beta E_h}{2}\Big)\Big]
+
 \frac{\Delta^2}{E_h^2} \frac{\cosh((\frac{\beta}{2}-\tau)E_h)}{\cosh(\beta E_h/2)}
 \Big].
 \label{gqkellelee}
 \end{align}
 The first term in the integral gives 
 \begin{align}
-\frac{g_n^2}{2N}
\sum_{\sigma\in\pm}
\frac{A_\sigma}{x_\sigma}
\frac{\partial^2}{\partial h^2} 
\frac{4h^2}{E_h^2}\Big[1-\tanh^2\Big(\frac{\beta E_h}{2}\Big)\Big]_{h=0}
=
-\frac{g_1^2}{\Delta}
\frac{4}{\Delta}\Big[1-\tanh^2\Big(\frac{\beta \Delta}{2}\Big)\Big],
\end{align}  
where we have used the sum rule \eqref{sumrulea}. The term proportional to the $\tanh^2$ cancels $\Pi_\text{stat}^\infty$, see Eq.~\eqref{psisttainftyat}. For the second term in the integral \eqref{gqkellelee} we compute
\begin{align}
&\int_0^\beta d\tau 
\frac{\cosh((\frac{\beta}{2}-\tau)x_\sigma)}{\sinh(\beta x_\sigma/2)}.
\frac{\cosh((\frac{\beta}{2}-\tau)E_h)}{\cosh(\beta E_h/2)}
=
\Big[\frac{\sinh(\frac{\beta}{2}(x_\sigma+E_h)}{x_\sigma+E_h}
+
\frac{\sinh(\frac{\beta}{2}(x_\sigma-E_h)}{x_\sigma-E_h}
\Big]\frac{1}{\sinh(\beta x_\sigma/2)\cosh(\beta E_h/2)}
\\
&
=
\frac{1}{E_h+x_\sigma}
\Big(1+\frac{\tanh(\frac{\beta}{2}E_h)}{\tanh(\frac{\beta}{2}x_\sigma)}\Big)
+
\frac{1}{x_\sigma-E_h}
\Big(1-\frac{\tanh(\frac{\beta}{2}E_h)}{\tanh(\frac{\beta}{2}x_\sigma)}\Big).
 \end{align}
 Finally, combining all terms
\begin{align}
\Pi_\text{stat}
&=
-\frac{4g_1^2}{\Delta^2}
-\frac{g_1^2}{E_h}
\sum_{\sigma\in\pm}
A_\sigma
\frac{\partial}{\partial E_h} 
\Big\{
 \frac{\Delta^2}{E_h^2} 
 \Big[
 \frac{1}{E_h+x_\sigma}
\Big(1+\frac{\tanh(\frac{\beta}{2}E_h)}{\tanh(\frac{\beta}{2}x_\sigma)}\Big)
+
\frac{1}{x_\sigma-E_h}
\Big(1-\frac{\tanh(\frac{\beta}{2}E_h)}{\tanh(\frac{\beta}{2}x_\sigma)}\Big)
 \Big]\Big\}_{h=0}.
 \label{gqkellelee888}
 \end{align}
Here, for taking the derivative, we note that for a function which depends only on $h^2$ we can replace $\partial_h^2 |_{h=0} = 2 \partial_{h^2} |_{h=0} = (4/E_h)\partial_{E_h}$. 
Equation~\eqref{gqkellelee888} is the exact result compared to exact diagonalization in the main text. 
It is illustrative to check the high-temperature limit, where $\tanh(\beta x/2)\to\beta x/2$:
\begin{align}
\Pi_\text{stat}(\beta\to 0)
&=
-\frac{4g_1^2}{\Delta^2}
-\frac{2g_1^2}{E_h}
\sum_{\sigma\in\pm}
\frac{A_\sigma}{x_\sigma}
\frac{\partial}{\partial E_h} 
 \frac{\Delta^2}{E_h^2} 
 \Big|_{h=0}
 =
-\frac{4g_1^2}{\Delta^2}
+\frac{4g_1^2}{\Delta^2}
\sum_{\sigma\in\pm}
\frac{A_\sigma}{x_\sigma}
=0.
\end{align}

\section{Interacting solid with all-to-all interaction}
\label{app:all2all}

\subsection{Imaginary-time action}
To discuss an interacting solid, we use the explicit model with a solid of $N$ molecules which are arranged in some fixed pattern in $3$ dimensions. Each molecule is described by a Hamiltonian $H_r$ (such as a two-level system). The bare matter Hamiltonian is then written as
\begin{equation}
\label{gd2ben}
	\hat{H}_{\rm mat}= \sum_{r} \hat H_r -\frac{\alpha}{2N}\sum_{r,r'} \hat p_{r}\hat p_{r}.
\end{equation}
The second term is an interaction between the molecules, where $\hat p$ is a dimensionless dipole operator. For the specific model, we consider an all-to-all interaction $f_{r,r'}=1/N$. This interaction will be assumed to favor a ferroelectric transition in the absence of the cavity, hence $\alpha>0$. The energy scale $\alpha$ will be chosen such that the transition is in the correct temperature range in the absence of the cavity.  The rest of the model, and the light-matter interaction will be as in Eq.~\eqref{Hgeneral}. 

After integrating out the light, we can write the action \eqref{effectiveca} in the form
\begin{align}
S_\text{eff}[c]&=\sum_{r}  S_\text{at}[c_r] - S_{V}[c],
\end{align}
where $S_\text{at}[c_r]$ is  is the action for the isolated molecule  (which depends only on the degrees of freedom related to site $r$), and the interaction term $S_V$ combines the direct interaction and the induced interaction 
\begin{align}
\label{sindgeheVV}
S_{V}=&\frac{1}{2N}\int_{0}^\beta\!\! d\tau \int_{0}^\beta \!\!d\tau'  P(\tau) 
V(\tau-\tau')
P(\tau'),
\end{align}
with $V(\tau)=\alpha\delta(\tau)-V_{\rm ind}(\tau)$, and the induced interaction \eqref{vinducedjs}. Note that we have chosen the sign such that $V$ is positive for small $g_1^2$.

Next, the bilinear term $S_V$ can again be decoupled with a single Hubbard Stratonovich transformation. Here we use the form
\begin{align} 
\label{actionphiwgwVV}
e^{S_{V}}=
\frac{1}{Z_V}
\int \mathcal{D}[\varphi] e^{
-\frac{1}{2}\int_0^\beta d\tau \int_0^\beta d\tau' \varphi(\tau)V^{-1}(\tau-\tau')\varphi(\tau')
-\int_0^\beta d\tau \sum_{r} \frac{\varphi(\tau)}{\sqrt{N}} p_{r}(\tau)},
\end{align}
to represent the induced interaction \eqref{sindgeheVV}, with an irrelevant constant $Z_V$. In contrast to Eq.~\eqref{actionphiwgw}, there is no factor $i$ in the coupling, because the action $S_V$ has opposite sign to $S_{\rm ind}$. This also reverses the sign in the exact relations
\begin{align}
\label{vgehjeveee01VV}
\chi(i\nu_m)
&=
-V(i\nu_m)^{-1}
+
V(i\nu_m)^{-1}
W(i\nu_m)
V(i\nu_m)^{-1},
\\
\label{vgehjeveeewwVV}
W(i\nu_m)
&=
V(i\nu_m)
+
V(i\nu_m)
\chi(i\nu_m)
V(i\nu_m).
\end{align}

Next we proceed with the solution of the problem, by integrating out the matter from the action. This follows the same steps as for the derivation of Eq.~\eqref{actioespddgw01}, with two differences: (i) Since the full interaction is decoupled and the remaining action $S_{\rm mat}$ describes isolated atoms, the susceptibilities $\chi_{\rm mat}^{(n)}$ can be replaced by the susceptibilities $\chi_\text{at}^{(n)}$ of the isolated atoms, and (ii), because of the factor $i$ in the coupling, integrating out matter yields a contribution $N\mathcal{G}_{\rm mat}[-i\varphi/\sqrt{N}]$ instead of $N\mathcal{G}_{\rm mat}[\varphi/\sqrt{N}]$, 
\begin{align} 
\label{actioespddgw02VV}
S&= \frac{1}{2}\int_0^\beta d\tau \int_0^\beta d\tau' \varphi(\tau) V^{-1}(\tau-\tau')\varphi(\tau')
+
N\mathcal{G}_{at}[-i\varphi/\sqrt{N}].
\end{align}
Because of the additional $i$ factor in $\mathcal{G}_\text{at}$, in the Taylor expansion the terms $n=2,6,10,...$ have a reversed sign. Hence, the total action in this case yields
\begin{align} 
\label{actioespddgw01VV}
S&= \frac{1}{2}\int_0^\beta d\tau \int_0^\beta d\tau' \varphi(\tau) W_0^{-1}(\tau-\tau')\varphi(\tau')
+
\frac{1}{N} S^{(4)} + \mathcal{O}(1/N^2),
\\
\label{S4VV}
S^{(4)}
&=\frac{1}{4!}\int_{1234} \chi^{(4)}_\text{at}(1,2,3,4) \varphi_1\varphi_2\varphi_3\varphi_4,
\end{align}
where 
\begin{align}
\label{WoVV}
W_0^{-1}=V^{-1}-\chi_\text{at}
\,\,\,\Rightarrow\,\,\,\,
W_0=\frac{V}{1-\chi_{at}V}.
\end{align}

\subsection{Mean-field theory}
\label{sec:int_MF}

The leading order in $1/N$ to \eqref{actioespddgw01VV} gives $W=W_0$, and with Eq.~\eqref{vgehjeveee01VV}
\begin{align} 
\chi_\text{mf}
&=
\frac{1}{V}\Big[
\frac{1}{1-\chi_\text{at}V}-1\Big]
=
\frac{\chi_{at}}{1-\chi_\text{at}V}.
\end{align}
This is again the mean-field description. In particular, for the static contribution $i\nu_m=0$ we have 
\begin{align} 
\chi_\text{mf}(0)
&=
\frac{\chi_\text{at}(0)}{1-\chi_\text{at}(0)\alpha},
\end{align}
which implies a divergence (second order phase transition) for the condition
\begin{align} 
1-\chi_\text{at}(0)\alpha=0.
\end{align}

\subsection{Static Hartree diagram}
\label{equ:int_hartree}
Similar as for the collective model, we can evaluate the Hartree diagram in the static limit, where
\begin{align}
\label{finalchialltoall}
\chi(0)=\frac{\chi_\text{at}(0)+\Pi(0)}{1-(\chi_\text{at}(0)+\Pi(0))\alpha},  \,\,\,\,\chi_\text{at}(0)=\frac{2}{\Delta} \tanh(\beta\Delta/2).
\end{align}
The evaluation for $\Pi(0)$ is then the same as for the Dicke model, and we can directly use Eq.~\eqref{pistat}
\begin{align}
\Pi_\text{stat} =
\frac{1}{2N}
\frac{\partial^2}{\partial h^2} 
\int_0^\beta d\tau \,W_{0}(\beta-\tau)\chi_{h}(\tau)\Big|_{h=0},
\end{align}
where $W_0$ is  now understood as Eq.~\eqref{WoVV}, and $\chi_h$ is given by Eq.~\eqref{chihat007}.
Again replacing $\partial_h^2 |_{h=0} = 2 \partial_{h^2} |_{h=0} $, 
\begin{align}
\frac{\partial^2}{\partial h^2} 
\chi_h(i\nu_m) |_{h=0}
&=
2 \partial_{h^2} 
\Big\{
 \frac{\Delta^2}{E_h}\tanh\Big(\frac{\beta E_h}{2}\Big) \frac{2}{E_h^2+\nu_m^2}
+
\beta\delta_{n,0}\frac{4h^2}{E_h^2}\Big[1-\tanh^2\Big(\frac{\beta E_h}{2}\Big)\Big]
\Big\}_{h=0}
\\
&=
\frac{4}{E_h} \frac{\partial}{\partial E_h} 
\Big\{
 \frac{\Delta^2}{E_h}\tanh\Big(\frac{\beta E_h}{2}\Big) \frac{2}{E_h^2+\nu_m^2}
\Big\}_{h=0}
+
\delta_{n,0}\frac{8\beta}{\Delta^2}\Big[1-\tanh^2\Big(\frac{\beta \Delta}{2}\Big)\Big]
\\
&=
-\frac{8}{\Delta} \tanh\Big(\frac{\beta \Delta}{2}\Big) 
\frac{\nu_m^2+3\Delta^2}{(\Delta^2+\nu_m^2)^2}
+
2 \beta
\cosh^{-2}\Big(\frac{\beta\Delta}{2}\Big) \frac{2}{\Delta^2+\nu_m^2}
+
\delta_{n,0}\frac{8\beta}{\Delta^2}\Big[1-\tanh^2\Big(\frac{\beta \Delta}{2}\Big)\Big].
\end{align} 

With this, we have 
\begin{align}
\Pi_\text{stat} =
\frac{1}{2N\beta}
\sum_{n} \,W_{0}(i\nu_m) \frac{\partial^2 \chi_{h}(i\nu_m)}{\partial h^2} \Big|_{h=0}.
\end{align}
The sum should be convergent because $|\chi_{h}(i\nu_m)| \sim n^{-2}$ for large $n$.

\section{Dielectric constant}
\label{sec:eps}
The dielectric constant of a material quantifies the response of its macroscopic polarization density $\vec{P}(\vec{r})$ to an external electric field $\vec{E}_{\rm ext}(\vec{r})$. For a homogeneous medium, it is defined by the relation
\begin{equation}
\vec{P}=(\varepsilon-1)\varepsilon_0\vec{E}_{\rm ext}.
\label{equ:mac_Pol}
\end{equation}
We want to derive an expression for $\varepsilon$ corresponding to our microscopic model. For that purpose, we first recall that the static susceptibility $\chi(0)$ determines the response of $\langle \hat{P} \rangle/N$ to an external field $h$ that couples to $\hat{P}$, i.e.,
\begin{equation}
	\frac{\langle \hat P \rangle}{N}=\chi(0)h.
	\label{equ:mic_Pol}
\end{equation}
Moreover, we can calculate the polarization density for a material of volume $V$ from the microscopic dipole moments using the equation
\begin{equation}
|\vec{P}|=\frac{\sum_{r}\langle \hat p_r \rangle ed}{V},
\end{equation}
and, thus, 
\begin{equation}
	|\vec{P}|=\frac{\langle \hat P \rangle}{V} ed.
	\label{equ:macPol_micPol}
\end{equation}

We can now substitute Eq.~\eqref{equ:mac_Pol} and \eqref{equ:mic_Pol} into Eq.~\eqref{equ:macPol_micPol}, which yields
\begin{equation}
(\varepsilon-1)\varepsilon_0|\vec{E}_{\rm ext}|=\chi(0)hed\frac{N}{V}
\label{equ:equal_mac_mic_P}
\end{equation}
for $\vec{E}_{\rm ext}\parallel \vec{P}$.
If we assume that the polarization density is homogeneous and the external electric field is uniform over the entire solid, the interaction energy in the macroscopic description is given by $-|\vec{P}||\vec{E}_{\rm ext}|V$. Comparing this to the external field term $-h\hat P$ in the microscopic model, we find that $|\vec{E}_{\rm ext}|ed=h$. With this, Eq.~\eqref{equ:equal_mac_mic_P} can be solved for $\varepsilon$, which yields
\begin{equation}
	\varepsilon=1+g_n^2\chi(0),
\end{equation} 
where the light-matter interaction strength is defined as $g_n^2=\frac{(ed)^2}{V\epsilon_0}N$.

\end{document}